\documentclass[useAMS,usenatbib,fleqn]{mn2e}
\usepackage[english]{babel}
\usepackage{fancyhdr}
\usepackage{amsfonts}
\usepackage{amsmath}
\usepackage{amssymb}
\usepackage[colorlinks,linkcolor=blue,citecolor=blue,urlcolor=blue,pdftitle={Properties of GAMA groups},pdfauthor={Viola}]{hyperref}
\usepackage{multicol}
\usepackage{layout}
\usepackage{graphicx}
\usepackage{subfigure}
\usepackage{threeparttable}
\usepackage{times}
\usepackage{morefloats}

\newif\ifAMStwofonts
\AMStwofontstrue

\newcommand{\be}{\begin{equation}}
\newcommand{\ee}{\end{equation}}
\newcommand{\ba}{\begin{eqnarray}}
\newcommand{\ea}{\end{eqnarray}}
\newcommand{\brr}{\begin{array}}
\newcommand{\err}{\end{array}}
\newcommand{\bc}{\begin{center}}
\newcommand{\ec}{\end{center}}
\newcommand{\hm}{\,h^{-1}{\rm Mpc}}
\newcommand{\hk}{\,h^{-1}{\rm kpc}}

\newcommand{\mincir}{\raise
  -2.truept\hbox{\rlap{\hbox{$\sim$}}\raise5.truept \hbox{$<$}\ }}
\newcommand{\magcir}{\raise
  -2.truept\hbox{\rlap{\hbox{$\sim$}}\raise5.truept \hbox{$>$}\ }}
\newcommand{\siml}{\raise
  -2.truept\hbox{\rlap{\hbox{$\sim$}}\raise5.truept \hbox{$<$}\ }}
\newcommand{\simg}{\raise
  -2.truept\hbox{\rlap{\hbox{$\sim$}}\raise5.truept \hbox{$>$}\ }}

\def\calD{{\cal D}}

\def\calR{{\cal R}}

\def\ba{{\bf a}}

\def\bc{{\bf c}}

\def\be{{\bf e}}

\newcommand{\pasa}{pasa}
\newcommand{\aj}{AJ}
\newcommand{\apj}{ApJ}
\newcommand{\apjl}{ApJ}
\newcommand{\apjs}{ApJS}
\newcommand{\aap}{A\&A}

\newcommand{\mnras}{MNRAS}
\newcommand{\nat}{Nature}

\newcommand{\prd}{Phys. Rev. D}

\newcommand{\jcap}{JCAP}

\newcommand{\physrep}{Phys. Rep.}
\newcommand {\apgt} {\ {\raise-.5ex\hbox{$\buildrel>\over\sim$}}\ }
\newcommand {\aplt} {\ {\raise-.5ex\hbox{$\buildrel<\over\sim$}}\ }

\defcitealias{kuijken/etal:2015}{K+15}
\defcitealias{Robotham11}{R+11}

\title[KiDS+GAMA: properties of galaxy groups]{Dark matter halo properties of GAMA galaxy groups from 100 square degrees of KiDS weak lensing data}
\author[Viola M. et al.] 
{M. Viola$^{1}$\thanks{viola@strw.leidenuniv.nl}, M. Cacciato$^{1}$, M. Brouwer$^{1}$, K. Kuijken$^{1}$, H. Hoekstra$^{1}$, P. Norberg$^2$, \and A.S.G. Robotham$^3$, E. van Uitert$^{4,5}$, M. Alpaslan$^{12}$, I.K. Baldry$^6$, A. Choi$^7$, \and J.T.A. de Jong$^{1}$, S.P. Driver$^{3,14}$, T. Erben$^{5}$, A. Grado$^9$, Alister W. Graham$^8$, C. Heymans$^7$, \and H. Hildebrandt$^{5}$, A.M. Hopkins$^{13}$, N. Irisarri$^1$, B. Joachimi$^{4}$, J. Loveday$^{11}$, \and L. Miller$^{10}$, R. Nakajima$^{5}$, P. Schneider$^{5}$, C. Sif\'on$^{1}$, G. Verdoes Kleijn$^{15}$ \\
\\
  $^1$Leiden Observatory, Leiden University, Niels Bohrweg 2, 2333 CA Leiden, The Netherlands.\\
  $^2$ICC, Department of Physics, Durham University, South Road, Durham DH1 3LE, UK.\\
  $^3$ICRAR, School of Physics, University of Western Australia, 35 Stirling Highway, Crawley, WA 6009, Australia.\\
  $^4$Department of Physics and Astronomy, University College London, Gower Street, London WC1E 6BT, UK.\\
  $^5$Argelander-Institut f{\"u}r Astronomie, Auf dem H{\"u}gel 71, D-53121 Bonn, Germany.\\
  $^6$Astrophysics Research Institute, Liverpool John Moores University, IC2, Liverpool Science Park, 146 Brownlow Hill, Liverpool L3 5RF, UK.\\
  $^7$Scottish Universities Physics Alliance, Institute for Astronomy, University of Edinburgh, Royal Observatory, Blackford Hill, Edinburgh, EH9 3HJ, UK.\\
  $^8$Centre for Astrophysics and Supercomputing, Swinburne University of Technology, Hawthorn, VIC, 3122, Australia.\\
  $^9$INAF-Osservatorio Astronomico di Capodimonte, Via Moiariello 16 -80131 Napoli Italy.\\
  $^{10}$Department of Physics, Oxford University, Keble Road, Oxford OX1 3RH.\\
  $^{11}$Astronomy Centre, University of Sussex, Falmer, Brighton BN1 9QH, UK.\\
  $^{12}$NASA Ames Research Centre, N232, Moffett Field, Mountain View, CA 94035, United States.\\
  $^{13}$Australian Astronomical Observatory, P.O. Box 915, North Ryde, NSW, Australia.\\
  $^{14}$Scottish Universities' Physics Alliance (SUPA), School of Physics and Astronomy, University of St Andrews, North Haugh, St Andrews, KY16 9SS, UK.\\
  $^{15}$Kapteyn Astronomical Institute, University of Groningen, P.O. Box 800, 9700 AV Groningen, The Netherlands.
}

\pubyear{2015}

\begin{document}
\label{firstpage}
\maketitle

\begin{abstract}
The Kilo-Degree Survey (KiDS) is an optical wide-field survey designed to map the matter distribution in the Universe using weak gravitational lensing. In this paper, we use these data to measure the density profiles and masses of a sample of $\sim \mathrm{1400}$ spectroscopically identified galaxy groups and clusters from the Galaxy And Mass Assembly (GAMA) survey.  We detect a highly significant signal (signal-to-noise-ratio $\sim$ 120), allowing us to study the properties of dark matter haloes over one and a half order of magnitude in mass, from $M\sim 10^{13}-10^{14.5} h^{-1}\mathrm{M_{\sun}}$.
We interpret the results for various subsamples of groups using a halo
model framework which accounts for the mis-centring of the Brightest Cluster Galaxy (used
as the tracer of the group centre) with respect to the centre of the group's dark matter halo. We find that the density profiles of the haloes are well described by an NFW profile with concentrations that agree with predictions from  numerical simulations.  In addition, we constrain scaling relations between the mass and a number of observable group properties. We find that the mass scales with the total r-band luminosity as a power-law with slope $1.16 \pm 0.13$ (1-sigma) and with the group velocity dispersion as a power-law with slope  $1.89 \pm 0.27$ (1-sigma). Finally, we demonstrate the potential of weak lensing studies of groups to discriminate between models of baryonic feedback  at group scales by comparing our results with the predictions from the Cosmo-OverWhelmingly Large Simulations (Cosmo-OWLS) project, ruling out models without AGN feedback. 
\end{abstract}


\section{Introduction}

Galaxy groups are the most common structures in the Universe, thus
representing the typical environment in which galaxies are found. In
fact, most galaxies are either part of a group or have been part of a
group at a certain point in time \citep{Eke04}.  However,
  group properties are not as well studied compared to those of more
  massive clusters of galaxies, or individual galaxies. This is
because groups are difficult to identify due to the small number of
(bright) members. Identifying groups requires a sufficiently deep \footnote{Fainter than the characteristic galaxy luminosity $L^*$ where the power-law form of the luminosity function cuts off}
spectroscopic survey with good spatial coverage, that is near 100\% complete. Even if a sample of groups is constructed, the
typically small number of members per group prevents reliable direct
dynamical mass estimates \citep{Carlberg01,Robotham11}. It is possible
to derive ensemble averaged properties \citep[e.g.,][]{More09}, but
the interpretation ultimately relies on either a careful comparison to
numerical simulations or an assumption of an underlying analytical model \citep[e.g.,][]{More11} .

For clusters of galaxies, the temperature and luminosity of the hot
X-ray emitting intracluster medium can be used to estimate masses
under the assumption of hydrostatic equilibrium. Simulations
\citep[e.g.,][]{Rasia06, Nagai07} and observations
\citep[e.g.,][]{Mahadavi13} indicate that the hydrostatic masses are
biased somewhat low, due to bulk motions and non-thermal pressure
support, but correlate well with the mass. In principle, it is
possible to apply this technique to galaxy groups; however, this is
observationally expensive given their faintness in X-rays, and consequently samples are generally small \citep[e.g.,][]{Sun09,Eckmiller11,Kettula13, Finoguenov15,Pearson15} and typically limited to the more massive systems.

Furthermore, given their lower masses and the corresponding lower gravitational binding energy, baryonic processes, such as feedback from star formation and active galactic nuclei (AGN) are expected to affect groups more than clusters \citep[e.g.,][]{McCarthy10,LeBrun14}. This may lead to increased biases in the hydrostatic mass estimates. The mass distribution in galaxy groups is also important for predictions of the observed matter power spectrum, and recent studies have highlighted that baryonic processes can lead to significant biases in cosmological parameter constraints from cosmic shear studies if left unaccounted for \citep[e.g.,][]{vDaalen11, Semboloni11, Semboloni13}. 

The group environment also plays an important role in determining the
observed properties of galaxies.  For example, there is increasing
evidence that star formation quenching happens in galaxy groups
\citep{Robotham13,Wetzel14}, due to ram pressure stripping, mergers, or AGN jets
in the centre of the halo \citep{Dubois13}. The properties of galaxies
and groups of galaxies correlate with properties of their host dark
matter halo \citep{Vale04,Moster10,Behroozi10,Moster13}, and the
details of those correlations depend on the baryonic processes taking
place inside the haloes \citep{LeBrun14}. Hence, characterisation of these
correlations is crucial to understand the effects of environment on galaxy evolution.

The study of galaxy groups is thus of great interest, but constraining models of galaxy evolution using galaxy groups requires both reliable and complete group catalogues over a relatively large part of the sky and unbiased measurements of their dark matter halo properties.  In the past decade, several large galaxy surveys have become available, and significant effort has been made to reliably identify bound structures and study their properties \citep{Eke04,Gerke05,Berlind06,Brough06,Knobel09}. In this paper, we use the group catalogue presented in \cite{Robotham11} (hereafter R+11) based on the three equatorial fields of the spectroscopic Galaxy And Mass Assembly survey \citep[herafter GAMA,][]{Driver11}.
For the reasons outlined above, determining group masses using
``traditional'' techniques is difficult. Fortunately, weak
gravitational lensing provides a direct way to probe the mass
distribution of galaxy groups \citep[e.g.,][]{Hoekstra01, Parker05, Leauthaud10}. It uses the tiny coherent distortions
in the shapes of background galaxies caused by the deflection of light
rays from foreground objects, in our case galaxy groups
\citep[e.g.,][]{Bartelmann01}. Those distortions are directly
proportional to the tidal field of the gravitational potential of the
foreground lenses, hence allowing us to infer the properties of their
dark matter haloes without assumptions about their dynamical
status. The typical distortion in the shape of a background object caused by
foreground galaxies is much smaller than its intrinsic ellipticity, preventing a precise mass determination for individual groups. Instead, we can only infer the ensemble averaged properties by averaging the shapes of many background galaxies around many foreground lenses, under the assumption that galaxies are randomly oriented in the Universe.

The measurement of the lensing signal involves accurate shape estimates, which in turn require deep, high quality imaging data. The shape measurements presented in this paper are obtained from the ongoing Kilo-Degree Survey (KiDS; \citealt{dejong/etal:2015}).  KiDS is an optical imaging survey with the OmegaCAM wide-field imager \citep{Kuijken11} on the VLT Survey Telescope \citep{Capaccioli11, deJong13} that will eventually cover 1500 square degrees of the sky in 4 bands ($ugri$). Crucially, the survey region of GAMA fully overlaps with KiDS. The depth of the KiDS data and its exquisite image quality are ideal to use weak gravitational lensing as a technique to measure halo properties of the GAMA groups, such as their masses. This is the main focus of this paper, one of a set of articles about the gravitational lensing analysis of the first and second KiDS data releases \citep{dejong/etal:2015}. Companion papers will present a detailed analysis of the properties of galaxies as a function of environment (van Uitert et al. in prep), the properties of satellite galaxies in groups \citep{sifon/etal:2015}, as well as a technical description of the lensing and photometric redshift measurements \citep[][K+15 hereafter]{kuijken/etal:2015}.

In the last decade, weak gravitational lensing analyses of large optical surveys have become a standard tool to measure average properties of dark matter haloes
\citep{Brainerd96,Fischer00,Hoekstra04,Sheldon04,Parker05,
Heymans06b, Mandelbaum06,Johnston07,Sheldon09,vanuitert/etal:2011,Leauthaud12,Choi12,Velander14, Coupon15, Hudson15}. However,
the interpretation of the stacked lensing signal of haloes with different properties is not trivial.  Haloes with different masses are stacked together, and a simple fit of the signal using some function describing an average halo profile, like a Navarro-Frenk-White profile \citep[][hereafter NFW]{NFW95} , can provide biased measurements.  A natural framework to describe the statistical weak lensing signal is the so-called halo model \citep{Cooray02,vdBosch13}. It provides a statistical description of the way observable galaxy properties correlate with the mass of dark matter haloes taking into account the halo mass function, the halo abundance and their large scale bias. 

The outline of this paper is as follows. In Section \ref{sec:statLensing}, we summarise the basics of weak lensing theory. We describe the data used in this work in Section \ref{sec:data}, and we summarise the halo model framework in Section \ref{sec:HaloModel}.  In Section \ref{sec:HaloProperies}, we present our lensing measurements of the GAMA galaxy groups, and in Section \ref{sec:ScalingRelation}, we derive scaling relations between lensing masses and optical properties of the groups. We conclude in Section \ref{sec:Conclusions}.

The relevant cosmological parameters entering in the calculation of
distances and in the halo model are taken from the Planck best fit
cosmology \citep{Planck13}: $\Omega_{\rm m}=0.315$, $\Omega_{\rm \Lambda}=0.685$, $\sigma_8 = 0.829$, $n_{\rm s}=0.9603$ and $\Omega_{\rm b} h^2=0.02205$.
Throughout the paper we use $M_{200}$ as a measure for the masses of the groups as defined by 200 times the mean density (and corresponding radius, noted as $R_{200}$).

\section{Statistical weak gravitational lensing}\label{sec:statLensing}

Gravitational lensing refers to the deflection of light rays from distant objects due to the presence of matter along the line-of-sight. Overdense regions imprint coherent tangential distortions (shear) in the shape of background objects (hereafter sources). Galaxies form and reside in dark matter haloes, and as such, they are biased tracers of overdense regions in the Universe. For this reason, one expects to find non-vanishing shear profiles around galaxies, with the strength of this signal being stronger for groups of galaxies as they inhabit more massive haloes. This effect is stronger in the proximity of the centre of the overdensity and becomes weaker at larger distances. 

Unfortunately, the coherent distortion induced by the host halo of a single galaxy (or group of galaxies) is too weak to be detected. We therefore rely on a statistical approach in which  many galaxies or groups that share similar observational properties are stacked together. Average halo properties (e.g. masses, density profiles) are then inferred from the resulting high signal-to noise shear measurements. This technique is commonly referred to as `galaxy-galaxy lensing', and it has become a standard approach for measuring masses of galaxies in a statistical sense. 

Given its statistical nature, galaxy-galaxy lensing can be viewed as a measurement of the cross-correlation of some baryonic tracer $\mathrm{\delta_{\rm g}}$ and the matter density field $\mathrm{\delta_{\rm m}}$:
\begin{equation}
\label{eq:Corr}
\xi_{\rm gm}(\textbf{r})=\langle \delta_{\rm g}(\textbf{x})\delta_{\rm m}(\textbf{x}+\textbf{r}) \rangle_{\textbf{x}}\, ,
\end{equation}
where $\textbf{r}$ is the three-dimensional comoving separation. The Equation above can be related to the projected matter surface density around galaxies via the Abel integral:
\begin{equation}
\label{eq:projectCorr}
\Sigma (R)= \bar{\rho}_{\rm m}\int_{0}^{\pi_{\rm s}} [1+\xi_{\rm gm}(\sqrt{R^2+\Pi^2})]\, \mathrm{d}\Pi\, ,
\end{equation}
where $R$ is the co-moving projected separation from the galaxy, $\pi_{\rm s}$ the position of the source galaxy,  $\mathrm{\bar{\rho}_{m}}$ is the mean density of the Universe
and $\Pi$ is the line-of-sight separation.\footnote{Here and
  throughout the paper we assume spherical symmetry. This assumption
  is justified in the context of this work since we measure the
  lensing signal from a stack of many different haloes with different
  shapes, which washes out any potential halo triaxiality.} Being sensitive to the density \emph{contrast}, the shear is actually a measure of the excess surface density (ESD hereafter):
\begin{equation}
\label{eq:ESD}
\Delta \Sigma (R)= {\bar \Sigma} (\le R)- \Sigma (R) \, , 
\end{equation}
where ${\bar \Sigma} (\le R)$ just follows from $\Sigma (R)$ via
\begin{equation}
\label{eq:SinR}{\bar \Sigma} (\le R) = \frac{2}{R^2} \int_0^R \Sigma (R') \, R'  \,  \mathrm{d}R' \, .
\end{equation}
The ESD can finally be related to the tangential shear distortion $\gamma_{\rm t}$ of background objects, which is the main lensing observable:
\begin{equation}
\Delta \Sigma (R)=\gamma_{\rm t} (R) \Sigma_{\rm cr} \, ,
\label{eq:deltaSigmaTheo}
\end{equation}
where
\begin{equation}
\Sigma_{\rm cr}=\frac{c^2}{4\pi G} \frac{D(z_{\rm s})}{D(z_{\rm l})D(z_{\rm l},z_{\rm s})} \, ,
\label{eq:SigmaCrit}
\end{equation}
is a geometrical factor accounting for the lensing efficiency. In the previous equation, $D(z_l)$ is the angular diameter distance to the lens, $D(z_{l},z_{s})$ the angular diameter distance between the lens and the source and $D(z_{s})$ the angular diameter distance to the source. 

In the limit of a single galaxy embedded in a halo of mass $M$, one can see that Equation \ref{eq:Corr} further simplifies because $\xi_{\rm gm}(\textbf{r})$ becomes the normalised matter overdensity profile around the centre of the galaxy. The stacking procedure builds upon this limiting case by performing a weighted average of such profiles accounting for the contribution from different haloes. This is best formulated in the context of the halo model of structure formation (see e.g. \citealt{Cooray02}, \citealt{vdBosch13}), and for this reason, we will embed the whole analysis in this framework (see Section~\ref{sec:HaloModel}).
In Section~\ref{sec:Measurements}, we describe how the ESD profile is measured.  

\section{DATA}
\label{sec:data}

\begin{figure*}
\centering{
\includegraphics[width=18cm, angle=0]{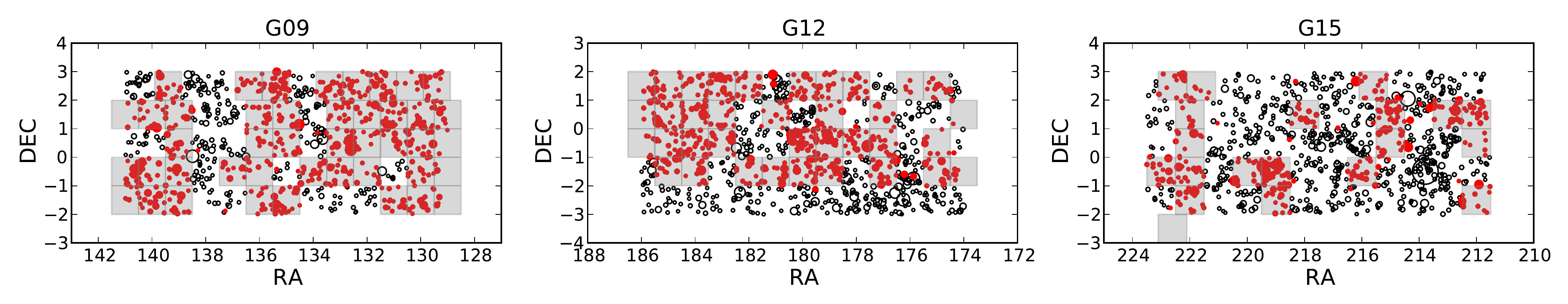}
\caption{KiDS-ESO-DR1/2 coverage of the three equatorial GAMA fields
  (G09, G12, G15). Each grey box corresponds to a single KiDS tile of 1 square degree. The black circles represent groups with $\mathrm{N_{fof}} \ge 5$ in the $\mathrm{G^3Cv7}$ catalogue \citepalias{Robotham11}.
The size of the dots is proportional to the group apparent richness. The filled red circles indicate the groups used in this analysis. These are all groups either inside a KiDS field or whose centre is separated less than 2 $h^{-1}\mathrm{Mpc}$ from the centre of the closest KiDS field. 
}}
\label{fig:footprint_DR2}
\end{figure*}

The data used in this paper are obtained from two surveys: the Kilo-Degree Survey (KiDS) and the Galaxy And Mass Assembly survey (GAMA). KiDS is an ongoing ESO optical imaging survey with the OmegaCAM wide-field imager on the VLT Survey Telescope \citep{deJong13}. When completed, it will cover two patches of the sky in four bands ($u,g,r,i$), one in the Northern galactic cap and one in the South, adding up to a total area of 1500 square degrees overlapping with the 2 degree Field Galaxy Redshift survey \citep[2dFGRS herafter,][]{Colless01}.  With rest-frame magnitude limits (5$\sigma$ in a 2" aperture) of 24.3, 25.1, 24.9, and 23.8 in the $u$, $g$, $r$, and $i$ bands, respectively, and better than 0.8 arcsec seeing in the $r$-band, KiDS was designed to create a combined data set that included good weak lensing shape measurements and good photometric redhifts. This enables a wide range of science including cosmic shear `tomography', galaxy-galaxy lensing and other weak lensing studies.

In this paper, we present initial weak lensing results based on observations of 100 KiDS tiles, which have been covered in all four optical bands and released to ESO as part of the first and second `KiDS-DR1/2' data releases to the ESO community, as described in \cite{dejong/etal:2015}. The effective area after removing masks and overlaps between tiles is 68.5 square degrees\footnote{A further 48 tiles from the KiDS-DR1/2, mostly in KiDS-South, were not used in this analysis since they do not overlap with GAMA.}.

In the equatorial region, the KiDS footprint overlaps with the footprint of the GAMA spectroscopic survey \citep{Baldry10,Robotham10,Driver11,Liske15},  carried out using the AAOmega multi-object spectrograph on the Anglo-Australian Telescope (AAT). The GAMA survey is highly complete down to petrosian $r$-band magnitude 19.8\footnote{The petrosian apparent magnitudes are measured from SDSS-DR7 and they include extinction corrections (Schlegel maps)},  and it covers $\sim \mathrm{180}$  square degrees in the equatorial region, which allows for the identification of a large number of galaxy groups.

Figure 1 shows the  KiDS-DR1/2 coverage of the G09, G12 and G15 GAMA fields. We also show the spatial distribution of the galaxy groups in the three GAMA fields (open black circles) and the selection of groups entering in this analysis (red closed circles).

\begin{table}
\label{tab:summary}
\caption{Summary of the area overlap of KiDS-DR1/2  in the three GAMA fields and the number of groups with at least 5 members used in this analysis. In parenthesis we quote the effective area, accounting for masks, used in this work.}
\begin{tabular}{lllll}
\hline
\multicolumn{1}{c}{GAMA field} & KiDS-DR1/2 overlap ($\mathrm{deg^2}$) & Number of groups &  &  \\
\hline
G09                            & 44.0 (28.5)          & 596             &  &  \\
G12                            & 36.0 (25.0)          & 509             &  &  \\
G15                            & 20.0 (15.0)          & 308             &  &  \\
\hline
\end{tabular}
\label{tab:gamafields}
\end{table}

\begin{figure}
\includegraphics[width=9cm, angle=0]{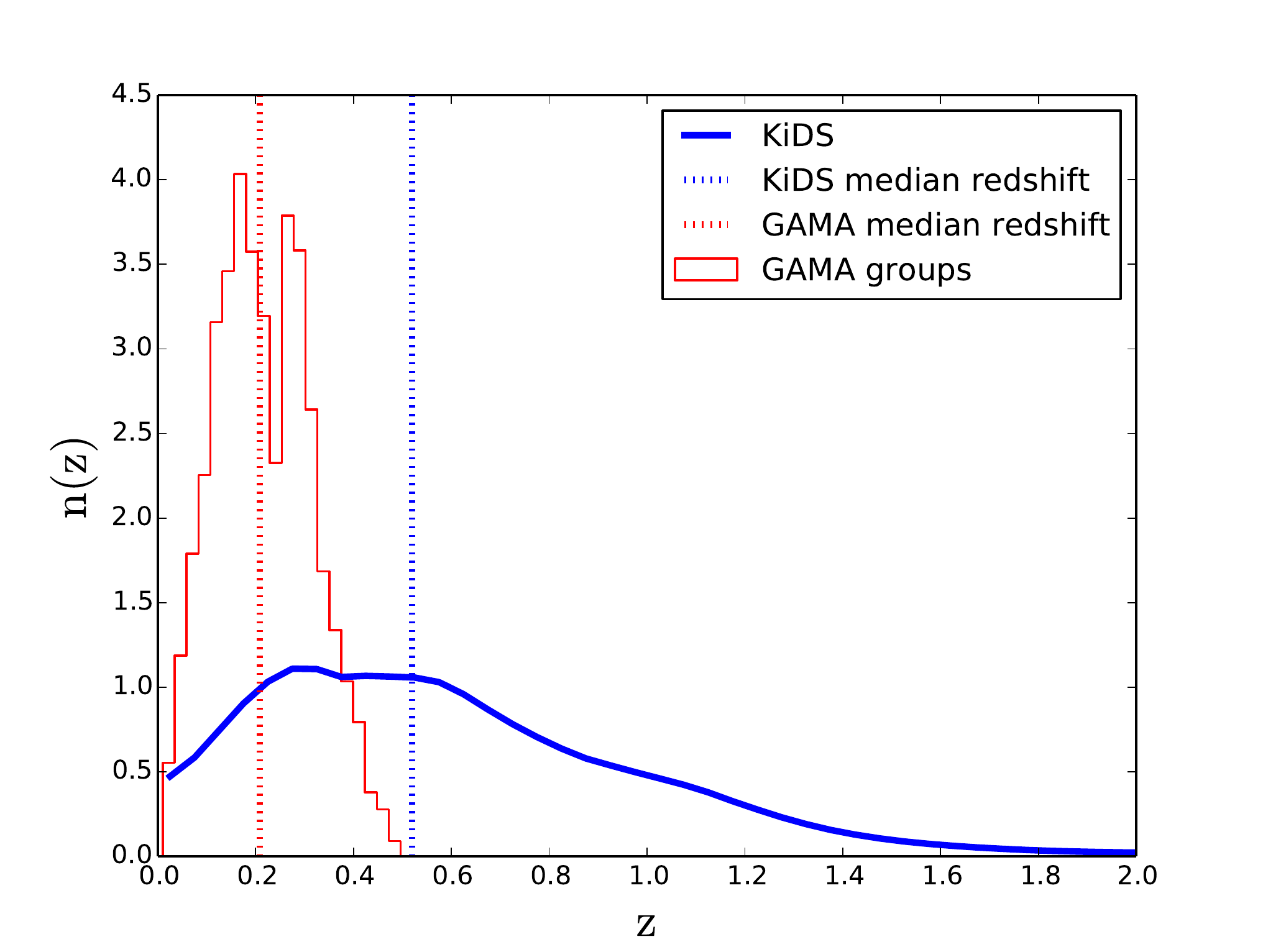}
\caption{Redshift distribution of the GAMA groups used in this analysis (red histogram) and the KiDS galaxies (blue lines). In the case of the GAMA groups, we use the spectroscopic redshift of the groups with at least 5 members \citepalias{Robotham11}, while for the KiDS galaxies the redshift distribution is computed as a weighted sum of the posterior photometric redshift distribution as provided by BPZ \citep{Benitez00}. The weight comes from \emph{lens}fit, used to measure the shape of the objects \citep{Miller07}. The two vertical lines show the median of the redshift distribution of the GAMA groups and of the KiDS sources. The two peaks in the redshift distribution of the GAMA groups are physical (and not caused by incompleteness), due to the clustering of galaxies in the GAMA equatorial fields.}
\label{fig:RedDis}
\end{figure}

Table~\ref{tab:gamafields} lists the overlap between KiDS-DR1/2 and GAMA and the total number of groups used in this analysis. Figure ~\ref{fig:RedDis} shows 
the redshift distribution of the GAMA groups used in this work and of the KiDS source galaxies, computed as a weighted sum of the posterior photometric redshift distribution as provided by BPZ \citep{Benitez00}. The weight comes from the \emph{lens}fit code, which is used to measure the shape of the objects \citep{Miller07} (see Sec. \ref{sec:shapeMeasure}). The median redshift of the GAMA groups is z=0.2, while the weighted median redshift of KiDS is 0.53. The multiple peaks in the redshift distribution of the KiDS sources result from degeneracies in the photometric redshift solution. This is dicussed further in \citetalias{kuijken/etal:2015}.
The different redshift distributions of the
two surveys are ideal for a weak lensing study of the GAMA groups using the KiDS galaxies as background sources.

\subsection{Lenses: GAMA Groups}
\label{sec:GAMA}
One of the main products of the GAMA survey is a group catalogue,
$\mathrm{G^{3}C}$ \citepalias{Robotham11}, of which we use the internal version 7. It
consists of 23,838 galaxy groups identified in the GAMA equatorial
regions (G09, G12, G15), with over 70,000 group members. It has been
constructed employing spatial and spectroscopic redshift information \citep{Baldry14}
of all the galaxies targeted by GAMA in the three equatorial
regions. The groups are found using a friends-of-friends algorithm, which links galaxies based on their projected and line-of-sight proximity. The choice of the linking length has been optimally
calibrated using mock data \citep{Robotham11,Merson13} based on the Millennium
simulation\footnote{$(\Omega_m,\Omega_b, \Omega_{\Lambda},h,\sigma_8, n_s)$=$(0.25,0.045,0.75,0.73,0.9,1.0)$} \citep{Springel05} and a semi-analytical galaxy formation model \citep{Bower06}. Running the final group selection
algorithm on the mock catalogues shows that groups with
at least 5 GAMA galaxies are less affected by interlopers and have sufficient members for a velocity dispersion estimate \citepalias{Robotham11}. For this reason we use only such groups in our analysis. 
This choice leaves us with 1413 groups, in  KiDS-DR1/2, 11\% of the full GAMA
group catalogue.

\begin{figure}
\includegraphics[width=9cm, angle=0]{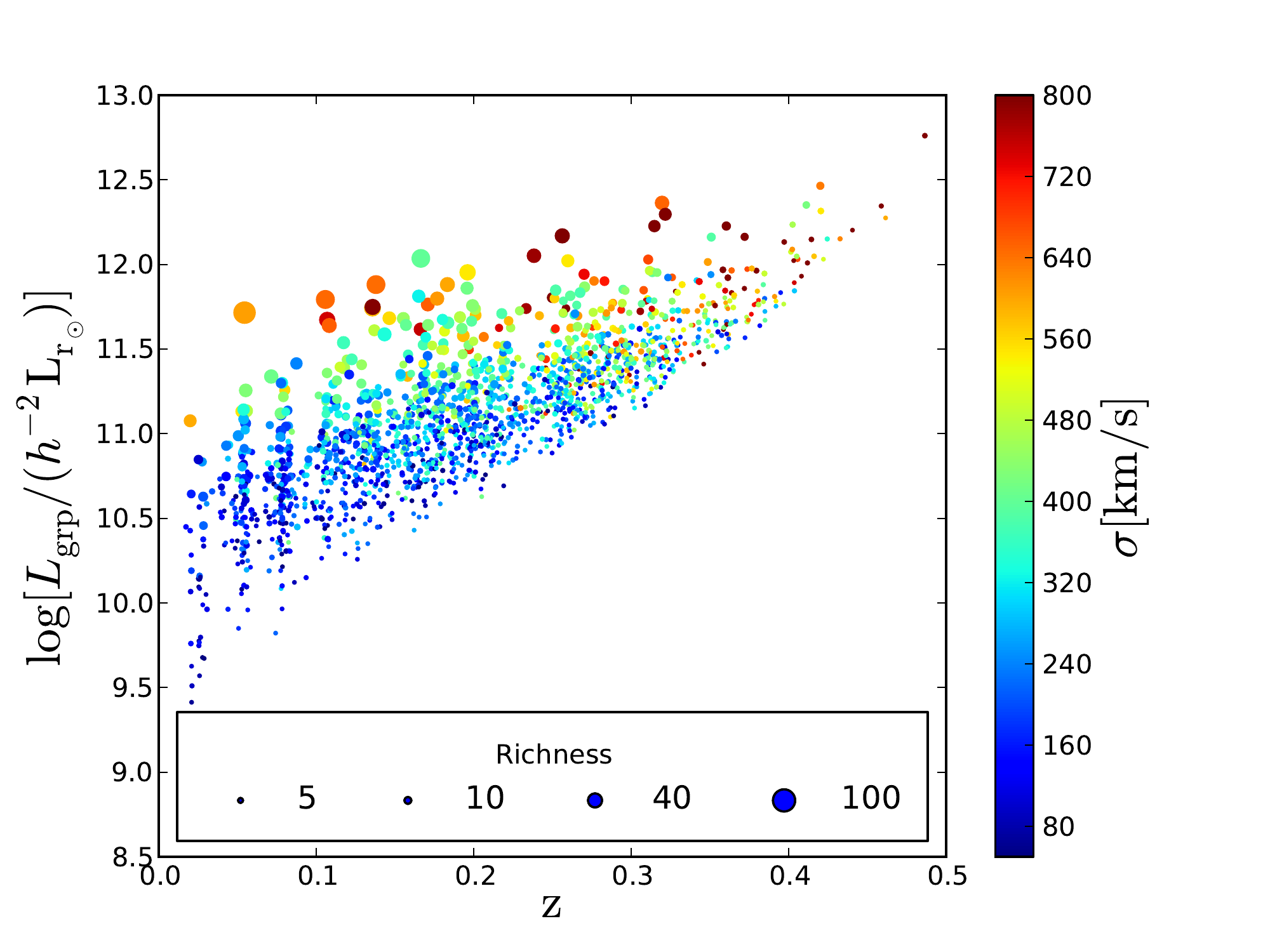}
\caption{Total group r-band luminosity as a function of the redshift of the group. The size of the points is proportional to the group apparent richness and the colour of the points indicates the group velocity dispersion corrected for velocity uncertainty. The shape of the distribution is typical of a flux limited survey.}
\label{fig:StatGroups}
\end{figure}

Figure~\ref{fig:StatGroups} shows the distribution of the total group r-band luminosity as a function of the redshift of the group, the group apparent richness, which is the number of members brighter that $r=19.8$, and the group velocity dispersion corrected for velocity uncertainty, for this subsample. These group r-band luminosity values are calculated by summing the r-band luminosity of all galaxies belonging to a group and targeted by GAMA and they also include an estimate of the contribution from faint galaxies below the GAMA flux limit, as discussed in \citetalias{Robotham11}. This correction is typically very small, a few percent at low redshift and a factor of a few at $z \sim 0.5$ since most of the luminosity comes from galaxies around $\mathrm{M^{\star}}-5\log h \sim -20.44$ \citep{Loveday12,Loveday15}, and most of the groups are sampled well below $\mathrm{M^{\star}}$. Note that all absolute magnitudes and luminosities used in the paper are k-corrected and evolution corrected at  redshift $z=0$ \citepalias{Robotham11}. The global k-correction used by \citetalias{Robotham11} is compatible with the median k-correction of the full GAMA \citep[][Fig.1 in the paper]{McNaught14}.

All the stellar masses used in this work are taken from \cite{Taylor11}, who fitted \cite{Bruzual03}  synthetic stellar spectra to the broadband SDSS photometry assuming a \cite{Chabrier03} IMF and a \cite{Calzetti00} dust law.

\subsection{Sources: KiDS galaxies}

We measure the gravitational lensing effect induced by the GAMA groups using galaxy images from KiDS. We refer to \citetalias{kuijken/etal:2015} for a detailed description of the pipelines used to measure shapes and photometric redshifts for those objects. We briefly summarise here the aspects of the data processing most relevant for this analysis. 

\subsubsection{Shape measurements}
\label{sec:shapeMeasure}
All of our lensing measurements are derived from the $r$-band
exposures in KiDS. This is the band with the highest image quality of the survey, as the
queue-scheduling at the telescope ensures that observations in this
filter are taken in the best seeing conditions. The images are
processed with the \textsc{THELI} pipeline, which has been optimized
for lensing applications \citep{Erben13}, and ellipticities for the
galaxies are derived using the \emph{lens}fit code
\citep{Miller07,Kitching08,Miller13}. \emph{lens}fit takes full
account of the point-spread function in the individual (dithered)
exposures and prior knowledge of the ellipticity and size distributions of faint galaxies, returning an ellipticity estimate for each galaxy as well as an inverse variance weight that is related to the uncertainty of the measurement. 

The average number density of galaxies with \emph{lens}fit weight $w$ larger than 0, and satisfying the photometric redshift cuts described in the next section, is 8.88 per square arcmin, corresponding to an effective number density:
\begin{equation}
n_{\mathrm{eff}}=\frac{\sigma^2_{\epsilon^{\rm s}}}{A} \sum_i w_i
\end{equation}
of 4.48 galaxies per square arcmin, where $A$ is the survey area and $\sigma^{2}_{\rm \epsilon^{\rm s}}=0.065$ is the intrinsic ellipticity variance. This is a measurement of the statistical power of the weak lensing data (see \cite{Chang13} and \citetalias{kuijken/etal:2015} for more details). 

It is well known that shape measurements for galaxies with low
signal-to-noise ratio and small sizes tend to be biased \citep[e.g.,][]{Melchior12, Refregier12,Miller13,Viola14}. 
This `noise-bias' stems from the non-linear transformations of the image pixels involved in the derivation of galaxy image shapes. It has the form of a multiplicative bias, and a calibration of the shape measurements is typically required in order to get an unbiased shear estimator. In this paper, we use the same calibration that was determined in \cite{Miller13}. This calibration depends on the signal-to-noise and the size of the objects and needs to be applied, in an average sense, to the recovered shear field. In addition to this multiplicative bias, shape measurements can also be affected by an additive bias caused by a non-perfect PSF deconvolution, centroid bias and pixel level detector effects. This bias can be empirically quantified and corrected  for directly from the data, using the residual average ellipticity over the survey area. More detail on these $\sim10$ per cent bias corrections can be found in \citetalias{kuijken/etal:2015}.

The analysis presented in this paper has been applied to four different ellipticity catalogues. Three of these catalogues were generated by rescaling all the ellipticity measurements by some factors unknown to the team and chosen by a collegue, Matthias Bartelmann\footnote{bartelmann@uni-heidelberg.de}, external to the collaboration. The amplitude of the rescaling has been chosen such that the cosmological parameters derived from a cosmic shear analysis using the four blind catalogues would not differ more than 10-$\sigma$, where sigma is the error from the Planck cosmological papers. We refer to this procedure as {\it blinding}, and we
have used it to mitigate confirmation bias in our data
analysis. The authors asked our external to {\it unblind}
the true shear catalogues only just before paper submission.
The authors were not allowed to change any of the results after the unblinding, without documenting those changes. Whilst the shear was blind, we did not blind measurements of group properties, such as their luminosity, or measurements of the source photometric redshifts.

\subsubsection{Photometric redshift measurements}

The observable lensing distortion depends on the distances to the lens and source (Equation~\ref{eq:SigmaCrit}). Redshifts to the lenses are known from the GAMA spectroscopy, but for the sources we need to resort to photometric redshifts derived from the KiDS-ESO-DR1/2 ugri images in the ESO data release. Processing and calibration of these images is done using the Astro-WISE environment \citep{mcfarland/etal:2013}, and flux and colour measurements use the `Gaussian Aperture and Photometry' (GAaP) technique designed to correct aperture photometry for seeing differences \citep{Kuijken08}. These colours form the basis of the photometric redshift estimates, obtained with BPZ \citep{Benitez00,Hildebrandt12}. After extensive tests, we reject galaxies whose photometric redshift posterior distribution $p(z)$ peaks outside the range [0.005,1.2] \citepalias{kuijken/etal:2015}. In what follows the $p(z)$ for each source is used in the calculation of distances, and in particular in the calculation of the critical surface density (see Equation \ref{eq:SigmaCrit}). \citetalias{kuijken/etal:2015} show that if the peak of each source's $p(z)$ had been used as the estimate of the redshift, the average value of $\Sigma_{\mathrm{cr}}$ and hence the average ESD would have been underestimated by $\sim$ 10 \%.

\subsection{Measurement of the stacked excess surface density profile}\label{sec:Measurements}

The shape measurement algorithm used in this work, \emph{lens}fit, provides measurements of the galaxy ellipticities ($\mathrm{\epsilon_{1}}$, $\mathrm{\epsilon_{2}}$) with respect to an equatorial coordinate system. 

For each source-lens pair we compute the tangential $\epsilon_{\rm t}$ and cross component $\epsilon_{\rm x}$ of the source's ellipticity around the position of the lens,
\begin{equation}
\begin{pmatrix} \epsilon_{\rm t} \\
\epsilon_{\rm x}
\end{pmatrix}= \begin{pmatrix} -\cos(2\phi) & -\sin(2\phi) \\
\phantom{-}\sin(2\phi) & -\cos(2\phi) 
\end{pmatrix} \begin{pmatrix} \epsilon_{1} \\
\epsilon_{2}
\end{pmatrix}
,
\end{equation}
where $\mathrm{\phi}$ is the position angle of the source with respect to the lens. 
The average of the tangential ellipticity of a large number of
galaxies in the same area of the sky is an unbiased estimate of the
shear. On the other hand, the average of the cross ellipticity over
many sources should average to zero. For this reason, the cross ellipticity is commonly used as an estimator of possible systematics in the measurements. 
Each lens-source pair is then assigned a weight
\begin{equation}
\tilde{w}_{\rm{ls}}=w_{\rm s}\tilde \Sigma_{\mathrm{cr}}^{-2} \, ,
\label{eq:weights}
\end{equation}
which is the product of the \emph{lens}fit weight $w_{s}$ assigned
to the given source ellipticity, and a geometric term $\tilde\Sigma_{\mathrm{cr}}$ which downweights lens-source pairs that are close in redshift and therefore less sensitive to lensing.
We compute the `effective critical surface density' for each pair from the spectroscopic redshift of the lens $z_l$ and the full posterior redshift distribution of the source, $p(z_s)$:
\begin{equation}
\tilde\Sigma_{\rm cr}^{-1}=\frac{4\pi G}{c^2}\int_{ z_l}^{\infty} \frac{D_l(z_l)D_{ls}(z_{l},z_{s})}{D_s(z_{s})}p(z_s)\mathrm{d}z_s \, .
\end{equation}
Finally, following Equation \ref{eq:deltaSigmaTheo}, we compute the ESD in bins of projected distance $R$ to the lenses:
\begin{equation}
\label{eq:ESDmeasured}
\Delta \Sigma (R)=\Bigg(\frac{\sum_{ls}\tilde{w}_{ls}\epsilon_{\rm t}\tilde\Sigma_{\rm cr}}{\sum_{ls}\tilde{w}_{ls}}\Bigg)\frac{1}{1+K(R)} \, ,
\end{equation}
where the sum is over all source-lens pairs in the distance bin, and
\begin{equation}
K(R)=\frac{\sum_{ls}\tilde{w}_{ls}m_{s}}{\sum_{ls}\tilde{w}_{ls}} \, ,
\end{equation}
is an average correction to the ESD profile that has to be applied to correct for the multiplicative noise bias $m$ in the \emph{lens}fit shear estimates. Typically, the value of the $K(R)$ correction is around 0.1, largely independent of the scale at which it is computed. 

\begin{figure}
\includegraphics[width=8cm, angle=0]{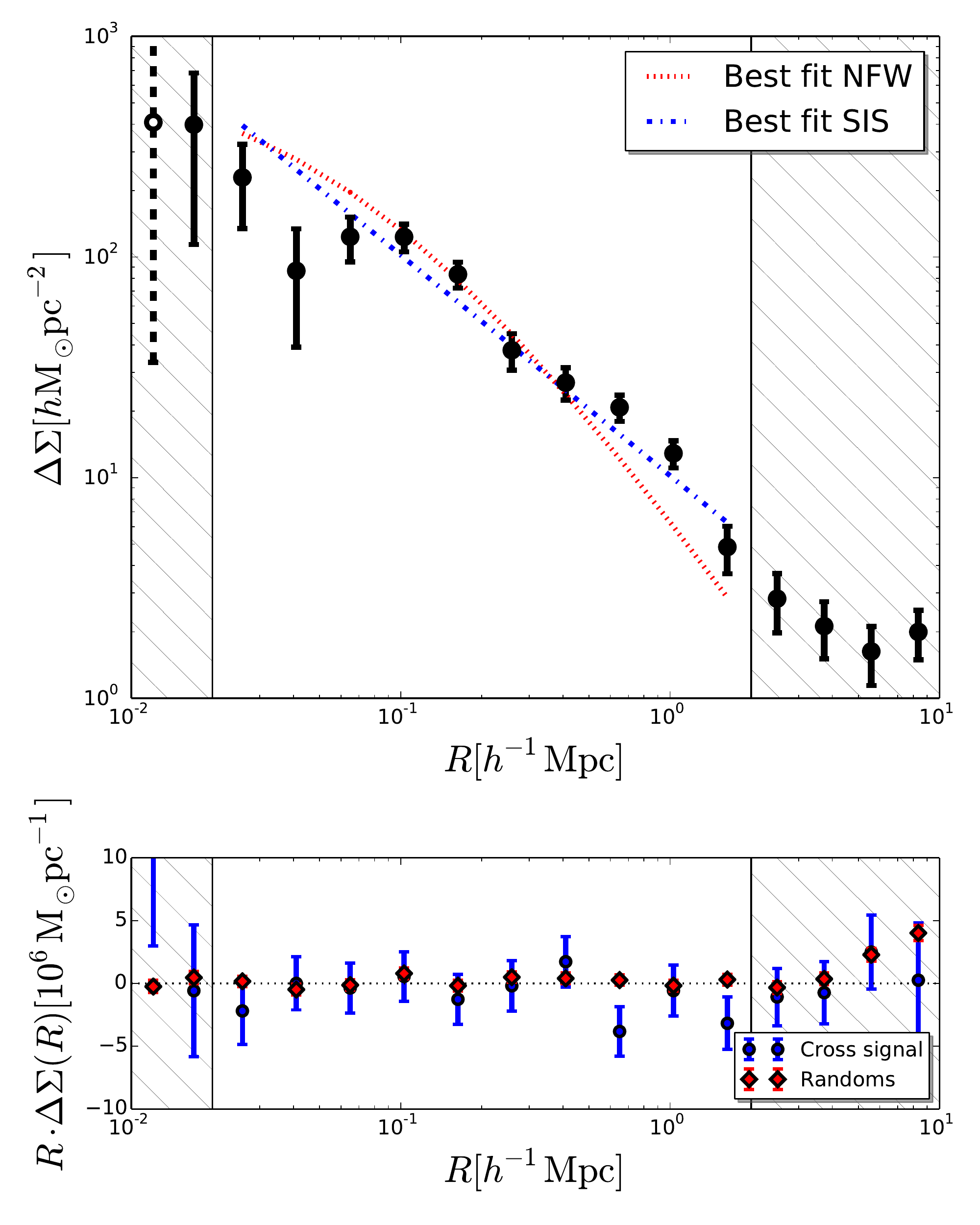}
\caption{\textit{Top panel:} ESD
  profile measured from a stack of all GAMA groups with at least 5
  members (black points). Here, we choose the Brightest Cluster Galaxy as the group
  centre. The open white circle with dashed error bars indicates a
  negative $\Delta\Sigma$. The dotted red line and the dash-dotted
  blue line show the best fits to the data of NFW \citep{NFW95} and singular
  isothermal sphere profiles, respectively. Neither of the single-parameter models provides a good fit to the data, highlighting that complex modelling of the signal is required. \textit{Bottom panel:} ESD profile, multiplied by $R$ to enhance features at large radii, measured from the cross-component of the ellipticities for these same groups (blue points) and measured around random points using the same redshift distribution of the groups (red points). We only use measurements at scales outside the dashed areas for the rest of the paper.}
\label{fig:SignalAllGroups}
\end{figure}

Figure~\ref{fig:SignalAllGroups} shows the stacked ESD profile for all
groups either inside a KiDS field or whose centre is separated by less than 2 $h^{-1}\mathrm{Mpc}$ from the centre of the closest KiDS field. It shows a highly significant detection of the lensing signal (signal-to-noise ratio $\sim$ 120). 
We note that the signal-to-noise is very poor at scales smaller than
$20\hk$. This is due to the fact that many objects close to the group
centres are blended, and \emph{lens}fit assigns them a vanishing
weight. We exclude those scales from any further analysis presented in this paper. 

For reference, we also show the best fit singular isothermal sphere (SIS) and NFW models to the stacked ESD signal. In the case of the NFW model, the halo concentration is fixed using the \citet{Duffy08} mass-concentration relation. Neither of the two single-parameter models provides a good fit to the data ($\chi^2_{red} >2.5$), highlighting how a more complex modelling of the signal is required (see Section \ref{sec:HaloModel}).

Figure ~\ref{fig:SignalAllGroups} also includes two tests for residual
systematic errors in the data: the cross-component of the signal and
the signal measured around random points in the KiDS tiles. On scales larger than $2\hm$, small but significant deviations are evident. We believe that one possible origin of the non-vanishing signal around random points at these scales is due to the incomplete azimuthal average of galaxy ellipticities, but we cannot exclude some large scale systematics in the shear data. The current patchy coverage of lensing data
complicates a detailed analysis and here we simply note that the
effect is small (less than 10 percent of the signal at $2\hm$) and
exclude data on scales larger than $2\hm$. Future analyses based on more uniform coverage of the GAMA area from the KiDS survey will need to address these potential issues. 

To summarise, in the rest of the paper we will use only projected distances in the range $(0.02-2)\hm$. Both the cross-component of the shear and the signal around random points are consistent with a null-detection over these scales.

\subsection{Statistical error estimate}
\label{sec:error}

In a stacking analysis with many foreground lenses, the ellipticity of
any source galaxy can contribute to the $\Delta\Sigma_i$ estimate in
multiple radial bins $i$ of different lenses. We summarize here how we compute the resulting covariances between the ESD estimates $\Delta\Sigma_i$ from the data.

We start from Equation~\ref{eq:ESDmeasured}, which gives the expression for  $\Delta\Sigma_i$.
For simplicity, we drop in what follows the noise bias correction factor $1+K(R)$ as it can be considered to have been absorbed in the effective critical density $\tilde\Sigma_{\rm cr}$.

We first rearrange the sum in Equation~\ref{eq:ESDmeasured} to separate the contributions from each source $s$, by summing first over all lenses $l$ that project within the radial bin $i$ from source $s$; for each source $s$ we denote this set of lenses as $i_s$. We can then rewrite Equation~\ref{eq:ESDmeasured}  as
\begin{equation}
	\Delta\Sigma_{i}=\frac{\sum_{s}w_{s}\left(\epsilon_{1s}C_{si}+\epsilon_{2s}S_{si}\right)}{\sum_{s}w_{s}Z_{si}} \, ,
\end{equation}
where $C$, $S$ and $Z$ are sums over the lenses
\begin{equation}
C_{si}=\sum_{l\in i_{s}}-\tilde\Sigma_{{\rm cr},ls}^{-1}\cos(2\phi_{ls})\, ,
\end{equation}
\begin{equation}
S_{si}=\sum_{l\in i_{s}}-\tilde\Sigma_{\rm {cr},ls}^{-1}\sin(2\phi_{ls})\, ,
\end{equation}
and
\begin{equation}
Z_{si}=\sum_{l\in i_{s}} \tilde\Sigma_{{\rm cr},ls}^{-2}\, .
\end{equation}
Since each $\epsilon_{ks}$ is an independent estimate of the shear field, where k=1,2, the ESD covariance between radial bins $i$ and $j$ can then be easily written as:
\begin{equation}\label{eq:covarianceRad}
	\mathbfss{Cov}_{ij}=\frac{\sum_{s}\sigma_{\epsilon}^{2}w_{s}^{2}\left(C_{si}C_{sj}+S_{si}S_{sj}\right)}{(\sum_{s}w_{s}Z_{si})(\sum_{s}w_{s}Z_{sj})} \, ,
\end{equation}
where $\sigma_{\epsilon}^{2}=0.078$ is the ellipticity dispersion weighted with the \emph{lens}fit weight, for one component of the ellipticity. We compute this number from the whole KiDS-ESO-DR1/2 area.

Equation \eqref{eq:covarianceRad} can be generalised to also compute the covariance between the ESD estimates for two different lens samples $m$ and $n$:
\begin{equation}
	\mathbfss{Cov}_{mnij}=\frac{\sum_{s}\sigma_{\epsilon}^{2}w_{s}^{2}\left(C_{si,m}C_{sj,n}+S_{si,m}S_{sj,n}\right)}{(\sum_{s}w_{s}Z_{si,m})(\sum_{s}w_{s}Z_{sj,n})} \, ,
\end{equation}
by restricting the sums for the $C$, $S$ and $Z$ terms to lenses in the relevant samples.

\begin{figure*}
\includegraphics[width=17cm, angle=0]{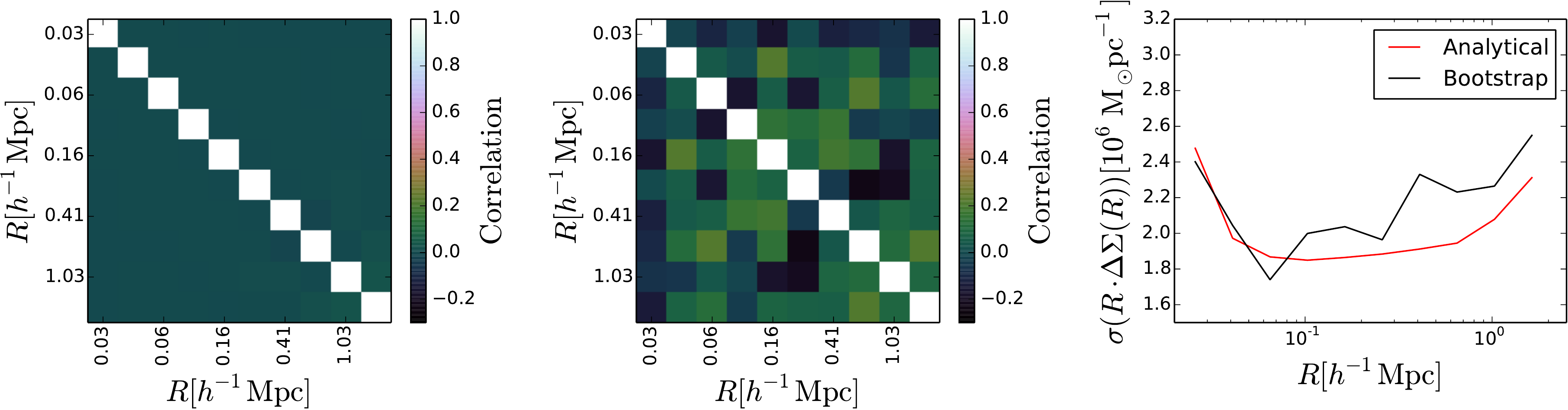}
\caption{\textit{Left panel:} ESD correlation matrix between different radial bins estimated from the data. This matrix accounts for shape noise and the effect of the mask and is computed as described in Section \ref{sec:error}. \textit{Middle panel:} ESD correlation matrix between different radial bins estimated using a bootstrap technique. It accounts for cosmic variance as well as shape noise. \textit{Right panel:} Comparison of the square root of the diagonal elements of the two covariance matrices as a function of distance from the group centre (here the BCG). Note the lower noise in the left-hand panel and the small but significant correlation between the largest-radial bins, which is a consequence of the many survey edges.}
\label{fig:CovMatrix}
\end{figure*}

We test the accuracy of the above calculation, which doesn't account for cosmic variance,  against the covariance matrix obtained via a bootstrapping technique. Specifically, we
bootstrap the signal measured in each of the 1-square degree KiDS
tiles. We limit the comparison to the case in which all groups are
stacked together\footnote{If the signal is split further into several
  bins according to some property of the group, we expect the relative
  contribution from cosmic variance compared to the contribution from
  shape noise to be even lower.} and compute the signal in 10
logarithmically spaced radial bins between $20\hk$ and $2\hm$. This
leads to an ESD covariance matrix with 55 independent entries, which
can be constrained by the 100 KiDS tiles used in this analysis. The
corresponding matrix is shown in Figure~\ref{fig:CovMatrix} together
with the correlation matrix obtained from Equation \ref{eq:covarianceRad}. The small but significant correlation between the largest-radial bins  is a consequence of the survey edges. We further show the diagonal errors obtained with the two methods, labelled Analytical and Bootstrap. Based on the work by \cite{Norberg09}, we might expect that the bootstrapping technique leads to somewhat larger error bars, although on larger scales this trend may be counteracted to some degree by the limited independence of our bootstrap regions. However, the conclusions of \cite{Norberg09} are based on an analysis of galaxy clustering, and a quantitative translation of their results to our galaxy-galaxy lensing measurements is not easy and beyond the scope of this work. The difference between the error estimates using these two independent methods is at most 10\% at scales larger than 300 $h^{-1}\mathrm{kpc}$.  
 
Based on the results of this test, we consider the covariance matrix
estimated from Equation \ref{eq:covarianceRad} to be a fair estimation of the
true covariance in the data, and we use it throughout the paper. 
In our likelihood analyses of various models for the
data (see next section), we account for the covariance between the
radial bins as well as between the different lens samples used to
compute the stacked signal.  
We note that future analyses with greater
statistical power, for example those based on the full KiDS and GAMA overlap,
and studies focusing on larger scales than those considered in this
analysis, will need to properly evaluate the full covariance matrix
that incorporates the cosmic variance contribution that is negligible in this work.

\section{Halo model}
\label{sec:HaloModel}

In this Section, we describe the halo model \citep[e.g.][]{Seljak00,Cooray02}, which we use to provide a physical interpretation of our data. 
We closely follow the methodology introduced in \cite{vdBosch13} and successfully applied to SDSS galaxy-galaxy lensing data in \cite{Cacciato13}.

This model provides the ideal framework to describe
the statistical weak lensing signal around galaxy groups.
It is based on two main assumptions: 
\begin{enumerate}
\item 
a statistical description of dark matter halo properties 
(i.e. their average density profile, their abundance and their large scale bias);
\item a statistical description of the way galaxies with different observable properties  populate dark matter haloes.
\end{enumerate}

As weak gravitational lensing is sensitive to the mass distribution
projected along the line-of-sight, the quantity of interest is the ESD
profile, defined in Equation \ref{eq:ESD}, which is related to the
galaxy-matter cross correlation via Equation \ref{eq:projectCorr}. Under the assumption that each galaxy group resides
in a dark matter halo, its average $\Delta\Sigma(R,z)$ profile can be computed using a statistical description of how galaxies are distributed over dark matter haloes of different mass and how these haloes cluster.
Specifically, it is fairly straightforward to obtain the two-point correlation function, $\xi_{\rm gm}(r,z)$, by Fourier transforming the galaxy-dark matter power-spectrum, $P_{\rm gm}(k,z) $, i.e.
\begin{eqnarray}\label{xiFTfromPK}
\xi_{\rm gm}(r,z) = {1 \over 2 \pi^2} \int_0^{\infty} P_{\rm gm}(k,z) 
\frac{\sin (kr)}{kr} \, k^2 \, {\rm d} k\,,
\end{eqnarray}
with $k$ the wavenumber, and the subscript `g' and `m' standing for `galaxy' and `matter'. 

In what follows, we will use the fact that, in Fourier space, 
the matter density profile of a halo of mass $M$ at a redshift $z$ can be described as
$ M \, {\tilde u}_{\rm h}(k|M,z)$,
where $M \equiv 4 \pi(200 {\bar \rho})R_{200}^3/3$, and 
 $\tilde{u}_{\rm h}(k|M)$ is the Fourier
transform of the normalized {\emph{dark}} matter density profile of a halo of mass $M$\footnote{We use $M_{200}$ masses for the groups throughout this paper, 
 i.e. as defined by 200 times the mean density (and corresponding radius, noted as $R_{200}$).}
We do not explicitly model the baryonic matter density profile \citep[][]{Fedeli14} because, on the scales of interest, its effect on the lensing signal can be 
approximated as that of a point mass (see Section ~\ref{sec:modelspecifics}).
Because the lensing signal is measured by stacking galaxy groups with observable property ${\cal O}_{\rm grp}$, on scales smaller than the typical extent of a group, we have $P_{\rm gm}(k,z)= P^{\rm 1h}_{\rm grp \, m}(k,z)$, where
\begin{equation}\label{eq:p1hgm}
P^{\rm 1h}_{\rm grp \, m}(k,z) = 
 \int {\cal P}(M|{\cal O}_{\rm grp}) \, {\cal H}_{\rm m}(k,M, z) \, {\rm d} M \, ,
\end{equation} 
and
\begin{eqnarray}\label{calHm0}
{\cal H}_{\rm m}(k,M, z) &\equiv & {M \over \bar{\rho}_{\rm m}} \, \tilde{u}_{\rm h}(k|M,z) \, ,
\end{eqnarray}
with
${\bar \rho}_{\rm m}$ the co-moving matter density of the Universe. Throughout the paper, the subscript `grp' stands for `galaxy group'.

The function ${\cal P}(M|{\cal O}_{\rm grp})$ is the probability that a group with observable property ${\cal O}_{\rm grp}$ resides in a halo of mass $M$. 
It reflects the halo occupation statistics and it can be written as: 
\begin{equation}
\label{eq:PP}
{\cal P}(M|{\cal O}_{\rm grp}) {\rm d} M =  {\cal H}_{\rm grp}(M,z)  \, n_{\rm h}(M,z) \, {\rm d} M \, . 
 \end{equation} 
Here, we have used
\begin{eqnarray}\label{calHg0}
{\cal H}_{\rm grp}(M,z) & \equiv & 
{\langle N \rangle_{{\cal O}_{\rm grp}}(M) 
\over \bar{n}_{\rm grp}({\cal O}_{\rm grp},z)} 
\,.
\end{eqnarray}
where, 
$\langle N \rangle_{{\cal O}_{\rm grp}}(M)$ is the average number of groups with observable property ${\cal O}_{\rm grp}$
that reside in a halo of mass $M$. 

Note that $n_{\rm h}(M,z)$ is the halo mass function (i.e. the number density of haloes as a function of their mass)
and we use the analytical function suggested in \cite{Tinker08} as a fit to a numerical N-body simulation.
Furthermore, the comoving number density of groups, ${\bar n}_{\rm grp}$, with the given observable property is defined as
\begin{equation}\label{eq:ngrp}
{\bar n}_{\rm grp}({\cal O}_{\rm grp},z) = \int \langle N \rangle_{{\cal O}_{\rm grp}}(M) \, n_{\rm h}(M,z) \, {\rm d} M \, .
 \end{equation} 

Note that in the expressions above we have assumed that we can
correctly identify the centre of the galaxy group halo (e.g., from the
position of the galaxy identified as the central 
in the GAMA group catalogue).
In Section \ref{sec:modelspecifics}, we generalize this expression to allow for possible {\it mis-centring} of the central galaxy.

Galaxy groups are not isolated, and on scales larger than the typical extent of a group,
one expects a non-vanishing contribution to the power spectrum due to the presence of other haloes surrounding the group.
This term is usually referred to as the two-halo term (as opposed to the one-halo term described in Equation \ref{eq:p1hgm}).
One thus has:
\begin{equation}
P_{\rm gm}(k) = P^{\rm 1h}_{\rm grp\, m}(k) + P^{\rm 2h}_{\rm grp\, m}(k) \,.
\end{equation} 
These terms can be written in
compact form as
\begin{equation}\label{P1h}
P^{\rm 1h}_{\rm grp\, m}(k,z) = \int {\cal H}_{\rm grp}(k,M,z) \, {\cal H}_{\rm m}(k,M,z) \, 
n_{\rm h}(M,z) \, {\rm d} M,
\end{equation}
\begin{eqnarray}\label{P2h}
\lefteqn{
P^{\rm 2h}_{\rm grp\, \rm m}(k,z) =
\int {\rm d} M_1 \, {\cal H}_{\rm grp}(k,M_1,z) \, n_{\rm h}(M_1,z) 
} 
\nonumber \\
& & \int {\rm d} M_2 \, {\cal H}_{\rm m}(k,M_2,z) \, n_{\rm h}(M_2,z) \,
Q(k|M_1,M_2,z)\,.
\end{eqnarray}

The quantity $Q(k|M_1,M_2,z)$ describes the power
spectrum of haloes of mass $M_1$ and $M_2$. 
In its simplest implementation\footnote{See, for example, \cite{vdBosch13} 
for a more refined description of this term.}, used throughout this paper, 
$Q(k|M_1,M_2,z)\equiv b_{\rm h}(M_1,z)b_{\rm h}(M_2,z)P^{\rm lin}(k,z)$, 
where  $b_{\rm h}(M,z)$ is the halo bias function
and $P^{\rm lin}(k,z)$ is the linear matter-matter power spectrum.
We note that, in the literature, there exist various fitting functions to describe 
the mass dependence  of the halo bias \citep[see for example][]{SMT,ST,Tinker10}. These functions 
may exhibit differences of up to $\sim 10$\% \citep[e.g.][]{Murray13}. However, a few points are worth a comment.

First, the use of the fitting function from \cite{Tinker10} is motivated by the use of a halo mass function 
calibrated over the same numerical simulation. Second,  the halo bias function enters 
in the galaxy-matter power spectrum only through the two-halo term and as part of an integral.
Thus, especially because we will fit the ESD profiles only up to $R=2 \, h^{-1}$Mpc, 
the uncertainty related to the halo bias function is much smaller than the statistical error
associated to the observed signal.

\subsection{Model specifics}
\label{sec:modelspecifics}

The halo occupation statistics of galaxy groups are defined via the function $\langle N \rangle_{{\cal O}_{\rm grp}}(M)$, 
the average number of groups (with a given observable property $\cal O_{\rm grp}$, such as a luminosity bin)
as a function of halo mass $M$. Since the occupation function of groups as a function of halo mass, $N_{\rm grp}(M)$, is either zero or unity,
one has that $\langle N \rangle_{{\cal O}_{\rm grp}}(M)$ is by construction confined between zero and unity.
We model $\langle N \rangle_{{\cal O}_{\rm grp}}(M)$ as a log-normal characterized by a mean, ${{\rm log}[{\tilde M}/(h^{-1} M_{\odot})]}$, 
and a scatter $\sigma_{\rm log{\tilde M}}$:
\begin{equation}\label{eq:phi_c}
\langle N \rangle_{{\cal O}_{\rm grp}}(M) \propto
{1\over {\sqrt{2\pi} \,  \sigma_{\log {\tilde M}} }} {\rm exp}
\left[- \frac{ (\log{M}-\log{{\tilde M}})^2}{2 \sigma^2_{\log {\tilde M}}}
\right]\,.
\end{equation}

We caution the reader against over-interpreting the physical meaning of this scatter;
this number mainly serves the purpose of assigning a distribution of masses around a mean value.

Ideally, for each stack of the group ESD (in bins of group luminosity or total stellar mass) we wish to determine both these parameters, but to keep the number of fitting parameters low we assume here that $\sigma_{{\rm log}\tilde M}$ is constant from bin to bin, with a flat prior  
$0.05 \le \sigma_{\rm log{\tilde M}} \le 1.5$. This prior does not have any statistical effect on the results and it only serves the purpose 
of avoiding numerical inaccuracies.
There is evidence for an increase in this parameter with central galaxy luminosity or stellar mass, \citep[e.g.][]{More09,More09b,More11}, 
but these increases are mild, and satellite kinematics
\citep[e.g.][]{More11} support the assumption that $\sigma_{\rm
  log{\tilde M}}$ is roughly constant on massive group scales (i.e. $\log
[M/(h^{-1}M_{\odot})] > 13.0$). We have verified that our assumption
has no impact on our results in terms of either accuracy or precision by allowing $\sigma_{\rm log{\tilde M}}$ to be different in each observable bin. 

For each given bin in an observable group property, one can define an effective mean halo mass, $\langle M \rangle$, as
\begin{eqnarray}
\label{eq:haloMasses}
\langle M \rangle_{{\cal O}_{\rm grp}} &\equiv& 
\int \, {\cal P}(M|{\cal O}_{\rm grp}) \, M \, {\rm d}M  \nonumber \\
&=& 
\frac{\int \langle N \rangle_{{\cal O}_{\rm grp}}(M) \, n_h(M,{\bar z})M \mathrm{d} M}
{{\bar n}_{\rm grp}({\cal O}_{\rm grp},{\bar z})} \, ,
\end{eqnarray}
where ${\bar z}$ is the mean redshift of the groups in the bin under consideration, and we have made use of Equation (\ref{eq:PP}) and (\ref{eq:ngrp}).
The effective mean halo mass, $\langle M \rangle_{{\cal O}_{\rm grp}} $, is therefore obtained as a weighted average where the weight is the multiplication of the halo 
occupation statistics and the halo mass function.

The dark matter density profile of a halo of mass $M$, $\rho_{\rm m}(r|M)$, is assumed to follow a NFW
functional form: 
\begin{equation}\label{NFW}
\rho_{\rm m}(r|M) = 
\frac{\overline{\delta}\,\overline{\rho}}{(r/r_s)(1+r/r_s)^{2}}\,,
\end{equation}
where $r_s$ is the scale radius and $\overline{\delta}$ is a
dimensionless amplitude  which can be  expressed in terms of  the halo
concentration parameter $c_{\rm m} \equiv R_{200}/r_s$ as
\begin{equation}\label{overdensity}
\overline{\delta} = {200 \over 3} \, {c_{\rm m}^{3} \over 
{\rm ln}(1+c_{\rm m}) - c_{\rm m}/(1+c_{\rm m})}\,.
\end{equation}
where the concentration parameter, $c_{\rm m}$, scales with halo mass.
Different studies in the literature have proposed somewhat different fitting functions \citep[e.g.][]{Bullock01b,Eke01,Maccio08,Duffy08,Klypin11,Prada12,Dutton14}
to describe the relation $c_{\rm m}(M,z)$.
Overall, these studies are in broad agreement but unfortunately have not converged 
to a robust unique prediction. 
Given that those fitting functions have been 
calibrated using numerical simulations with very different configurations
(most notably different mass resolutions and cosmologies), 
it remains unclear how to properly account for the above mentioned discrepancies.
As these fitting functions all predict a weak mass dependence, 
we decide to adopt an effective concentration-halo mass relation that has the
mass and redshift dependence proposed in \cite{Duffy08}
but with a rescalable normalization:
\begin{eqnarray}
\label{eq:cNorm}
&c_{\rm m}^{\rm eff}(M_{200},z) =  f_{\rm c}\times c_{\rm m}^{\rm Duffy}(M_{200},z)& \nonumber \\
&= f_{\rm c}\times  10.14 \left(\frac{M_{200}}{2 \times 10^{12}}\right)^{-0.081} (1+z)^{-1.01}&\, .
\end{eqnarray} 
Note that at $z=0.25$, one has $c_{\rm m}^{\rm Duffy} \approx 5$
for halo masses with $\log{[M/(h^{-1}M_{\odot})]} \approx 14.3$.
We leave $f_{\rm c}$ free to vary within a flat uninformative prior $0.2 \le f_{\rm c} \le 5$.

The innermost part of a halo is arguably the site where a `central' galaxy resides.
The baryons that constitute the galaxy may be distributed according to different profiles
depending on the physical state \citep[for example, exponential discs for stars and $\beta$-profiles for hot gas, see][]{Fedeli14}.
The lensing signal due to these different configurations could in principle be modelled to a certain level of sophistication
\citep[see][]{Kobayashi15}. 
However, at the smallest scales of interest here \footnote{We fit the data in the range $0.02<R/(h^{-1}\mathrm{Mpc})<2.0$},
those distributions might as well be accounted for by simply assuming a point mass, $M_{\rm P}$. 
In the interest of simplicity, we assume that the stellar mass of the 
brightest cluster galaxy ($M^{\rm BCG}_{\star}$ \citep{Taylor11} 
is a reliable proxy for the amount of mass in the innermost 
part of the halo. Specifically, we assume that 
\begin{equation}
M_{\rm P} = A_{\rm P} M_{\star}^{\rm BCG} \, ,
\end{equation}  
where $A_{\rm P} $ is a free parameter, within a flat prior between 0.5 and 5.

The adopted definition of centre may well differ from the true minimum of the gravitational potential well. Such a mis-centring of the `central' galaxy  
is in fact seen in galaxy groups (see e.g. \citealt{Skibba11} and references therein). 
\cite{George12} offer further independent support of such a mis-centring, finding that massive central galaxies trace the centre of mass to  less than 75 kpc/$h$.
 
We model this mis-centring in a statistical manner (see also \citealt{Oguri11}, \citealt{Miyatake13}, \citealt{More14} and references therein). Specifically, we assume that 
the degree of mis-centring of the groups in three dimensions, $\Delta(M,z)$, is proportional to the halo scale radius $r_s$, a function of halo mass and redshift, and parametrize the probability that a `central' galaxy is
mis-centred as $p_{\rm off}$. This gives
\begin{equation}
{\cal H}_{\rm grp}(k,M,{\bar z}) = 
{\langle N \rangle_{{\cal O}_{\rm grp}}(M) \, \over \bar{n}_{\rm grp}({\bar z})}  \, 
(1-p_{\rm off} + p_{\rm off}\times
{\mathrm e}^{[-0.5 k^2 (r_s {\cal R}_{\rm off})^2]}
)
\, .
\end{equation}
Setting either $p_{\rm off}$ or ${\cal R}_{\rm off}$ to zero implies that there is {\it de facto} no offset. 
We treat the two as free parameters in Section ~\ref{sec:HaloProperies}. The parameter $p_{\rm off}$, being a probability, is bound between zero and unity.
We apply a flat uniform prior to ${\cal R}_{\rm off} \in [0,1.5]$. We note that this prior is very conservative,
as according to  \cite{George12}  and \cite{Skibba11} 
the mis-centring is expected to be smaller than the scale radius of a group, for which ${\cal R}_{\rm off}=1$.   

In summary, the model parameter vector, is defined as 
$\lambda = ({\rm log {\tilde M}}_i, \sigma_{\rm log{\tilde M}},f_{\rm c},A_{\rm P}, p_{\rm off}, \calR_{\rm off})$ 
where $i = 1...N_{\rm bins}$. Throughout the paper, we bin group observable properties in 6 bins. This leads to a 11 parameter model.  
We use Bayesian inference techniques to determine the posterior probability distribution $P(\lambda | \calD)$ of the model parameters 
given the data, $\calD$.  According to Bayes' theorem,
\begin{equation}
P(\lambda | \calD) \propto P(\calD | \lambda) \, P(\lambda)\propto \exp\left[\frac{-\chi^2(\lambda)}{2} \right]\, P(\lambda)
 \,,
\end{equation}
where $P(\calD | \lambda)$ is the likelihood of the data given the model parameters,assumed to be gaussian, and $P(\lambda)$ is the prior probability of these parameters.
Here, 
\begin{equation}
\chi^2(\lambda)= 
[
\widetilde{\Delta\Sigma}_{k,j} - \Delta\Sigma_{k,j}
]^T 
(\mathbfss{C}^{-1})_{kk',jj'}
[
\widetilde{\Delta\Sigma}_{k',j'} - \Delta\Sigma_{k',j'}
]
\,,
\end{equation}
where $\Delta\Sigma_{k,j}$ is the $j$'th radial bin of the observed stacked ESD for the groups in bin $k$, and $\widetilde{\Delta\Sigma}_{k,j}$ is the corresponding model prediction. $\mathbfss{C}$ is the full covariance matrix for the measurements, computed as detailed in Section ~\ref{sec:error}.

We sample the posterior distribution of our model parameters given the data using a Markov Chain Monte-Carlo (MCMC). In particular, we use\footnote{A python implementation of this sampling method is available via the \textsc{MontePython} code thanks to the contribution by Surhud More.} a proposal 
distribution that is a multi-variate Gaussian whose covariance is computed via a Fisher analysis run during the burn-in phase of the chain, set to 5000 model evaluations.

\section{Density profile of galaxy groups}\label{sec:HaloProperies}

We measure the ESD signal around each GAMA group with at least 5 members in 10 logarithmically-spaced radial bins in the range 20 $h^{-1}\mathrm{kpc}$ to 2 $h^{-1}\mathrm{Mpc}$.
We first assign errors to those measurements by propagating the shape noise on the tangential shear measurement in each radial bin.
We divide the groups into 6 bins according to a given observable
property, such as their velocity dispersion, total r-band luminosity, apparent richness or r-band luminosity fraction of the BCG. 
Bin limits are chosen to make the signal-to-noise of the ESD roughly the same in each bin. Once the bin limits are defined, we compute the data covariance between radial bins and between group bins as outlined in Section \ref{sec:error}. We summarise the bin-limits, the number of groups in each bin, the mean redshift of the bin and the mean stellar mass of the BCG in Table \ref{tab:binSummary} for the four observables considered in this work.

\begin{table*}
\caption{Summary of the bin limits used to compute the stacked ESD signal, the number of groups in each bin, the mean redshift of the groups in each bin and the mean stellar mass of the BCG.}
\label{tab:binSummary}
\resizebox{\linewidth}{!}{%
\begin{tabular}{lllll}
\hline
$\texttt{Observable}$ & $\texttt{Bin limits}$ &$\texttt{Number of lenses}$ & $\texttt{Mean redshift}$& $\log(\langle M^{BCG}_{\star}[h^{-2}\mathrm{M_{\odot}}] \rangle)$\\
\hline
$\log[\mathrm{L_{grp}}/(h^{-2}\mathrm{L_{\odot}})]$& $(9.4,10.9,11.1,11.3,11.5,11.7,12.7)$ & $(540,259,178,233,142,66)$ & $(0.13,0.20,0.23,0.26,0.30,0.35)$& $(11.00, 11.23,11.29,11.37,11.47,11.70)$\\\
$\mathrm{\sigma/(s^{-1}km)}$&$(0,225,325,375,466,610,1500)$ & $(501,359,124,198,147,89)$ & $(0.15,0.19,0.21,0.23,0.26,0.31)$ & $(11.05, 11.20,11.30,11.36,11.41,11.64)$\\
$\mathrm{N_{fof}}$ & $(5,6,7,8,11,19,73)$ & $(481,261,170,239,181,86)$ & $(0.21,0.21,0.21,0.19,0.18,0.16)$& $(11.17, 11.23,11.29,11.29,11.35,11.45)$\\
$\mathrm{L_{BCG}/L_{grp}}$ & $(1.0, 0.35, 0.25, 0.18, 0.13, 0.08, 0)$ & $(346,252,296,227,200,97)$ & $(0.10, 0.16, 0.20, 0.25, 0.29, 0.34)$& $(11.16, 11.19,11.22,11.29,11.36,11.53)$\\
\hline
\end{tabular}}
\end{table*}

The typical signal-to-noise ratio in each of the 6 luminosity bins is of order $\sim$ 20-25. This is comparable to the signal-to-noise ratio reported by \cite{Sheldon09} for a weak lensing analysis of $\sim$ 130000 MaxBCG clusters using SDSS imaging, once we restrict the comparison to a similar luminosity range.

\begin{figure}
\includegraphics[width=8cm, angle=0]{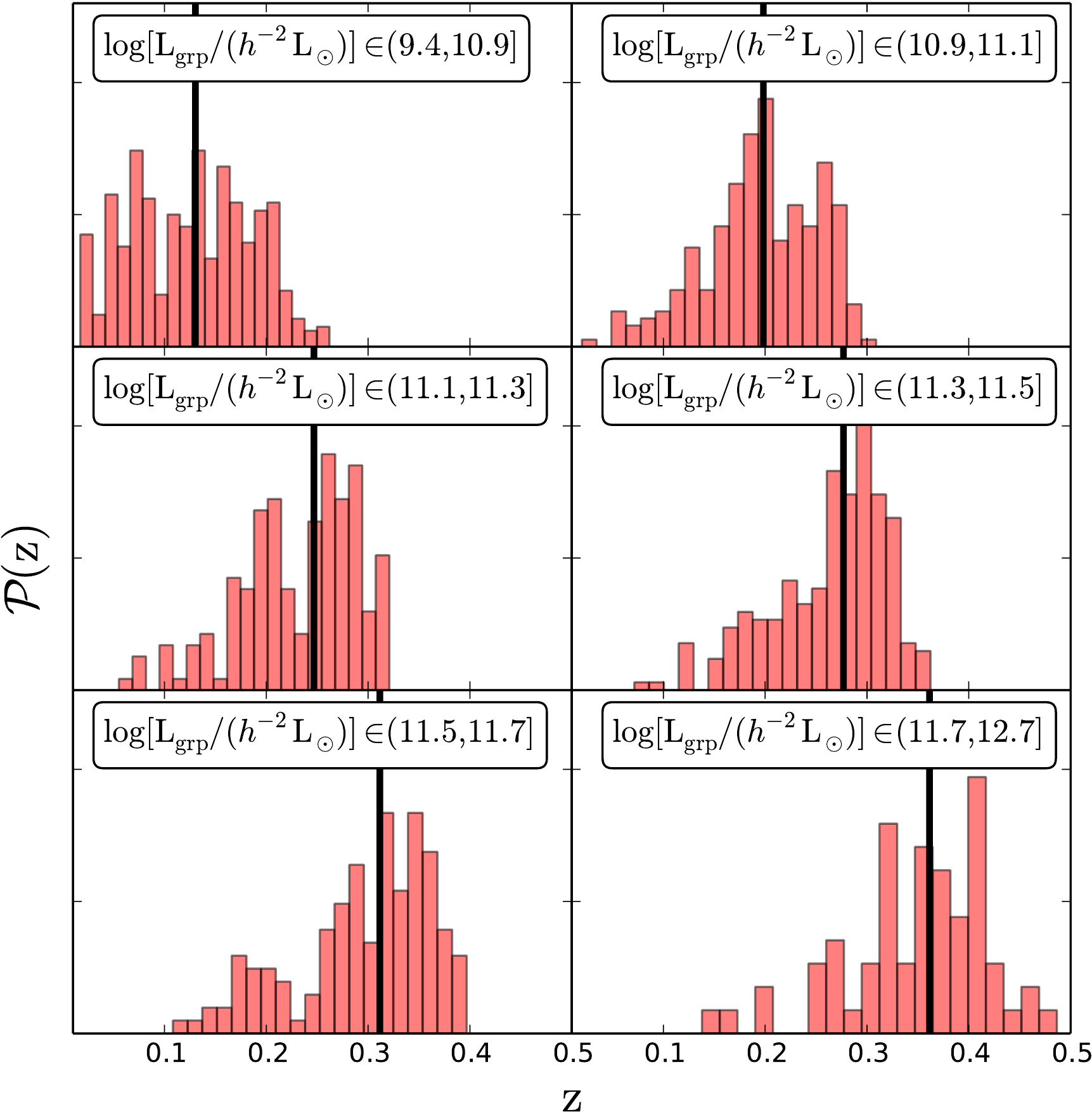}
\caption{Redshift distributions of the GAMA groups used in this paper in the six r-band luminosity bins. The group luminosity increases from left to right and from top to bottom. The solid vertical black lines indicate the median of the distributions.}
\label{fig:RedDisLumBin}
\end{figure}

We jointly fit the signal in the 6 bins using the halo model described
in Section~\ref{sec:HaloModel}. Since GAMA is a flux limited survey, the
redshift distributions of the groups in the six luminosity bins are
different, as shown in Figure \ref{fig:RedDisLumBin}. When we fit the
halo model to the data, we calculate the power spectra and mass
function  (Equations \ref{eq:p1hgm}-\ref{P2h}) using the median of the redshift distribution in each bin.

For each observable property, we run 5 independent chains with
different initial conditions. We evaluate the convergence of the MCMC
by means of a Gelman Rubin test \citep{Gelman92}, and we impose
$\mathrm{R} < 1.03$, where the $\mathrm{R}$-metric is defined as the
ratio of the variance of a parameter in the single chains to the
variance of that parameter in an "\textit{{\"u}ber-chain}", obtained by combining 5 chains.

\subsection{Matter density profiles of group-scale haloes}
 
We first test whether the ESD measurements themselves support the halo model assumption that the group density profile can be described in terms of a mis-centred NFW profile with a contribution from a point-mass at small scales, and what constraints can be put on the model parameters.
In the interest of being concise, we only present the results derived by
binning the groups according to their total r-band luminosity (see Section \ref{sec:data}), as statistically equivalent results are obtained when the groups are binned according to their velocity dispersion, apparent richness or r-band luminosity fraction of the BCG. The binning by other observables will become important in the study of scaling relation presented in Section \ref{sec:ScalingRelation}.   

One needs to  define the centre of the halo before stacking the ESD profiles of the groups.  
Following \citetalias{Robotham11}, we have three choices for the group centre: the centre of light (\texttt{Cen}), the Brightest Cluster Galaxy (\texttt{BCG}) and the brightest galaxy left after iteratively removing the most distant galaxies from the group centre of light (\texttt{IterCen}).  Throughout the paper, unless stated otherwise, we use the \texttt{BCG} as the definition of the centre, as it is a common choice in the literature. We investigate the effect of using the other two definitions of the group centre in Section \ref{sec:misCenter} and in  Appendix \ref{sec:OtherCentre}.

\begin{figure*}
\includegraphics[width=17.5cm, angle=0]{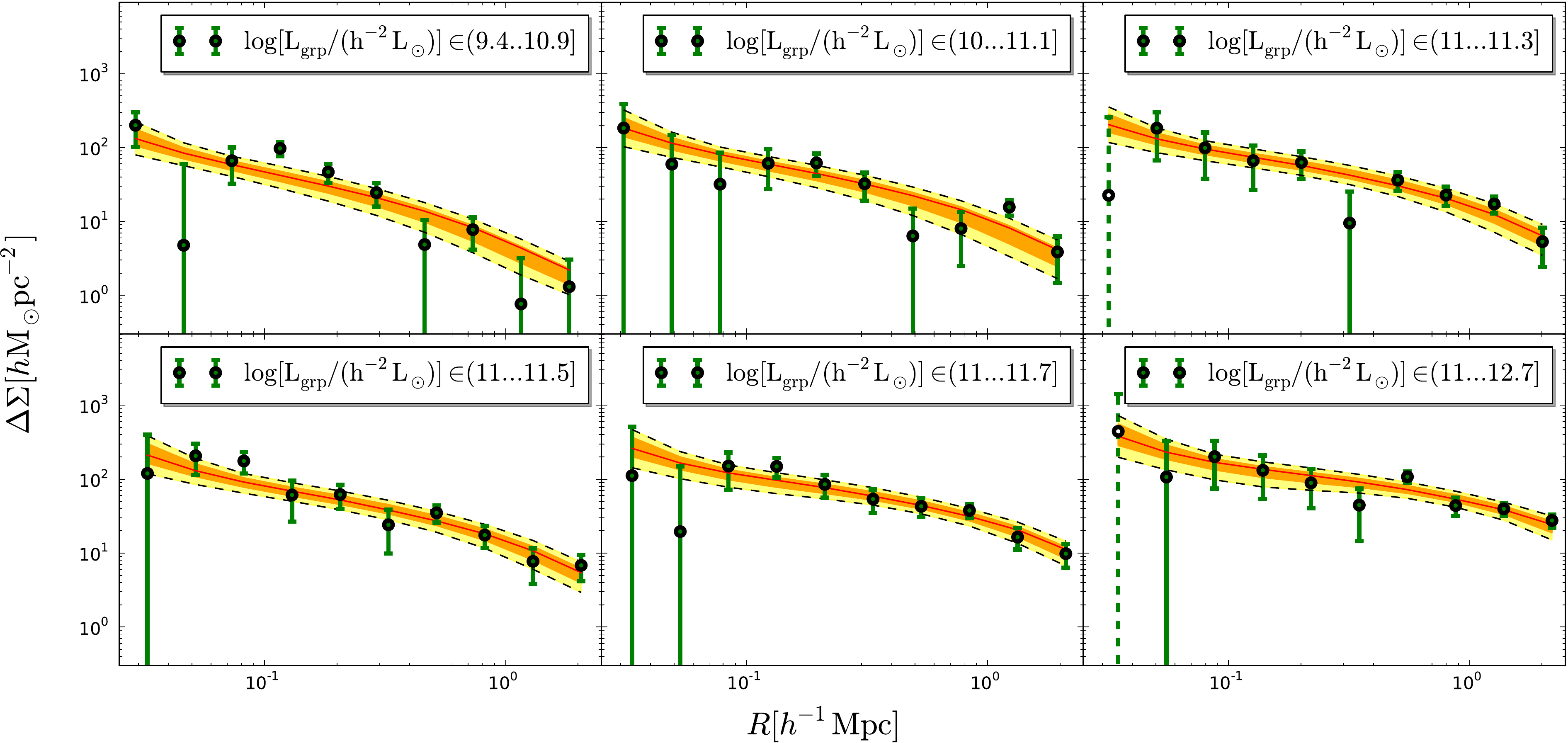}
\caption{Stacked ESD profile measured around the groups $\mathrm{BCG}$ of the 6  group luminosity 
bins as a function of distance from the group centre. The group r-band luminosity increases from left to right and from top to bottom. 
The stacking of the signal has been done using only groups with $\mathrm{N_{fof}} \ge 5$. The error bars on the stacked signal are computed 
as detailed in section \ref{sec:error} and we use dashed bars in the case of negative values of the ESD. The orange and yellow bands represent the 68 and 95 percentile of the model around the median, 
while the red line shows the best fit model.}
\label{fig:GroupSignalFiducial}
\end{figure*}

Figure \ref{fig:GroupSignalFiducial} shows the stacked ESD profiles (green points with error bars) for the 6 bins in total r-band luminosity. Note that the error bars are the square root of the diagonal elements of the full covariance matrix, and we use dashed bars in the case of negative values of the ESD. The ESD profiles have high signal-to-noise throughout the range in total luminosity and in spatial scales. Red lines indicate the best-fit model, whereas orange and yellow bands indicate the 68 and 95\% confidence interval. The model describes the data well with a reduced $\mathrm{\chi_{red}^{2}}=1.10$, 49 d.o.f, over the full scale range, for all the luminosity bins. 
This justifies our assumption that the ESD profile can be accurately modelled as a weighted stack of mis-centred NFW density profiles with a contribution from a point mass at the centre.

\begin{figure}
\includegraphics[width=8.0cm, angle=0]{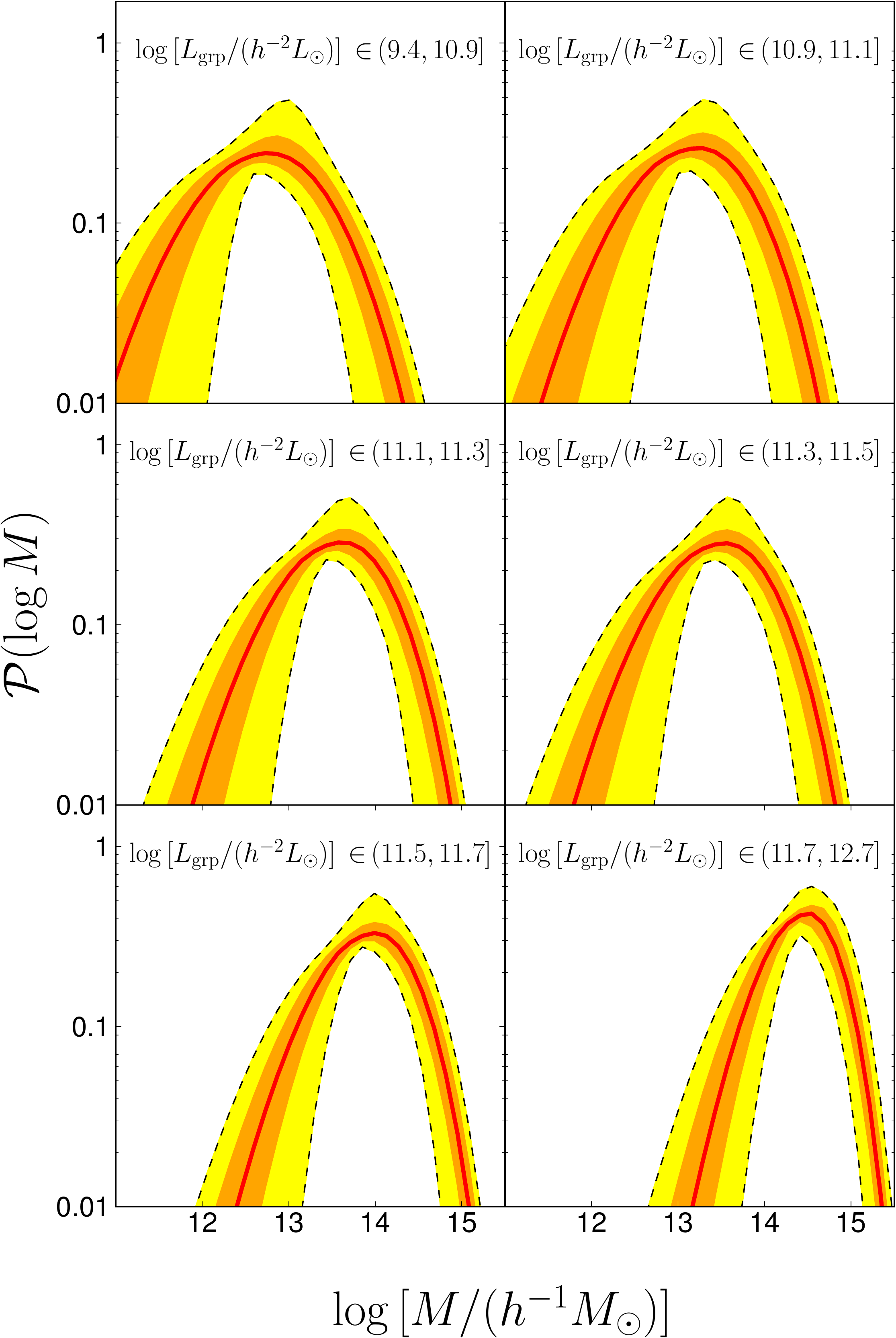}
\caption{Probability that a group with a given r-band luminosity resides in a halo of mass $M$. The red lines show the median distribution, while the orange and the yellow contours show the 68 and 95 percentile around the median.}
\label{fig:HODpost}
\end{figure}

The main results of this analysis can be summarised as follows (68\% percent confidence limits quoted throughout): 
\begin{itemize} 
\item For each r-band luminosity bin, we derive the probability that a group
  with that luminosity resides in a halo of mass $M$ (see
  Equation \ref{eq:PP}). We show the median of the probability distribution
  for the 6 bins in Figure \ref{fig:HODpost}. We constrain the scatter in the  mass at a fixed total r-band luminosity to be $\sigma_{\rm log{\tilde M}}=0.74^{+0.09}_{-0.16}$. This sets the width of the log-normal distribution describing the halo occupation statistics. We remind the reader that $\sigma_{\rm log \tilde{M}}$ is the width of the distribution in halo masses at given total luminosity of the groups and it is {\it not} the scatter in luminosity (or stellar mass) at a fixed halo mass that is often quoted in the literature and that 
one would expect to be considerably smaller \citep[e.g.][]{Yang09, Cacciato09, More11,Leauthaud12}. 
This hampers the possibility of a one-to-one comparison with most studies in the literature. However, we note that \cite{vdBosch07} and \cite{More11} reported values 
of the scatter in halo mass at fixed luminosity that are as high as 0.7 at the bright end. Furthermore, \cite{More15} reported a value of $0.79^{+0.41}_{-0.39}$ for the width of the low mass end distribution of the halo occupation statistics of massive CMASS galaxies. Given the non-negligible differences between the 
actual role of this parameter in all these studies, we find this level of agreement satisfactory.
\item  For each luminosity bin, a mean halo mass is inferred with a typical uncertainty on the mean of $\sim$ 0.12 dex.
\item The relative normalisation of the concentration-halo mass relation (see Equation \ref{eq:cNorm}) is constrained to be $f_{\rm c}=0.84^{+0.42}_{-0.23}$, in agreement with the nominal value based on \cite{Duffy08}.
\item The probability of having an off-centred BCG is $\mathrm{p_{off}} < 0.97$ (2-sigma upper limit), whereas the average amount of mis-centring in terms of the halo scale radius, $\cal R_{\rm off}$, is unconstrained within the prior.
\item  The amount of mass at the centre of the stack which contributes as a point mass to the ESD profiles is 
constrained to be $M_{\rm PM}=\mathrm{A_{\rm PM}}\, \langle M_{\star}^{BCG}\rangle =2.06^{+1.19}_{-0.99} \, \langle M_{\star}^{BCG}\rangle$. 
\end{itemize}

\begin{figure*}
\centering{
\includegraphics[width=18cm, angle=0]{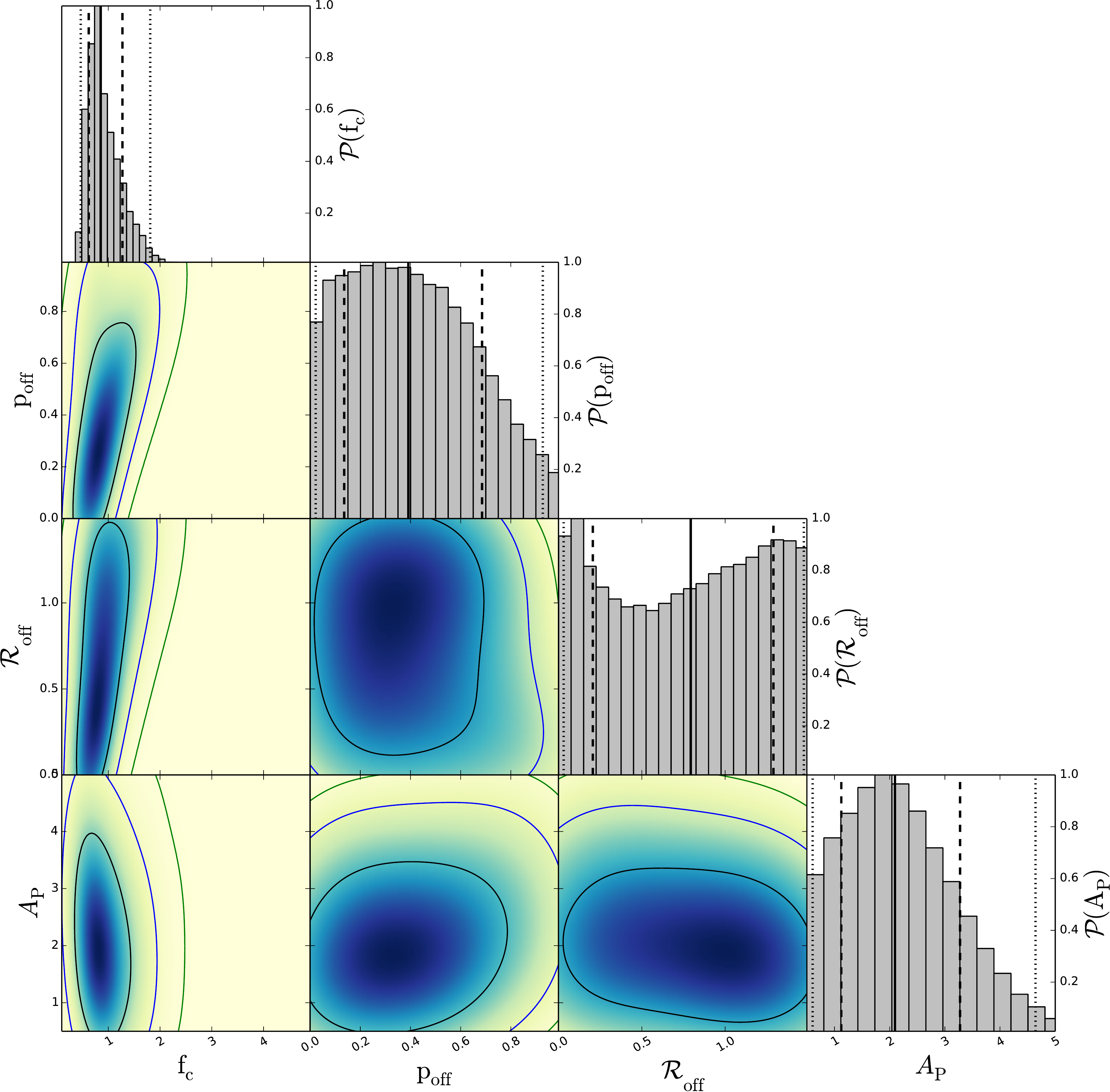}
\caption{Posterior distribution of the normalisation of the mass-concentration relation $f_{\rm c}$, of the mis-centring parameters $\mathrm{p_{off}}$ and $\mathrm{{\cal R}_{\rm off}}$ and of the amplitude of the point mass $\mathrm{A_{P}}$. The contours indicate the 1, 2, 3 sigma confidence regions. The dashed vertical lines and the dotted vertical lines correspond respectively to the 1 and 2 sigma marginalised confidence limits. These are the constraints from a joint halo model fit of the ESD signal in the 6 luminosity bins  using \texttt{BCG} as the group centre. The range in each panels reflect the priors used for the different parameters.}
\label{fig:AllparamBCG}}
\end{figure*}

Figure \ref{fig:AllparamBCG} shows the posterior distributions of the halo model parameters and their mutual degeneracies. Table \ref{tab:summaryParam} and \ref{tab:summaryParamExtra} list the median values of the parameters of interest with errors derived from the 16th and 84th percentiles of the posterior distribution. 
We discuss the constraints on the model parameters in further detail in the remainder of this section.

\subsubsection{Masses of dark matter haloes}

The dark matter halo masses of the galaxy groups that host the stacked galaxy groups analysed in this work span one and a half orders of magnitude with $M \in [10^{13}..10^{14.5}] h^{-1}\mathrm{M_{\odot}}$. 
Since our ESD profiles extend to large radii, our $2\hm$ cut-off is larger than $R_{200}$ over this full mass range, these mass measurements are robust and direct as they do not require any extrapolation. The uncertainties on the masses are obtained after marginalising over the other model parameters. Typically these errors are 15\% larger than what would be derived by fitting an NFW profile to the same data,  ignoring the scatter in mass inside each luminosity bins. Note that a simple NFW fit to the data in the 6 luminosity bins, with fixed concentration \citep{Duffy08} would also lead to a bias in the inferred masses of approximately 25\%. 

\begin{figure}
\includegraphics[width=8.5cm, angle=0]{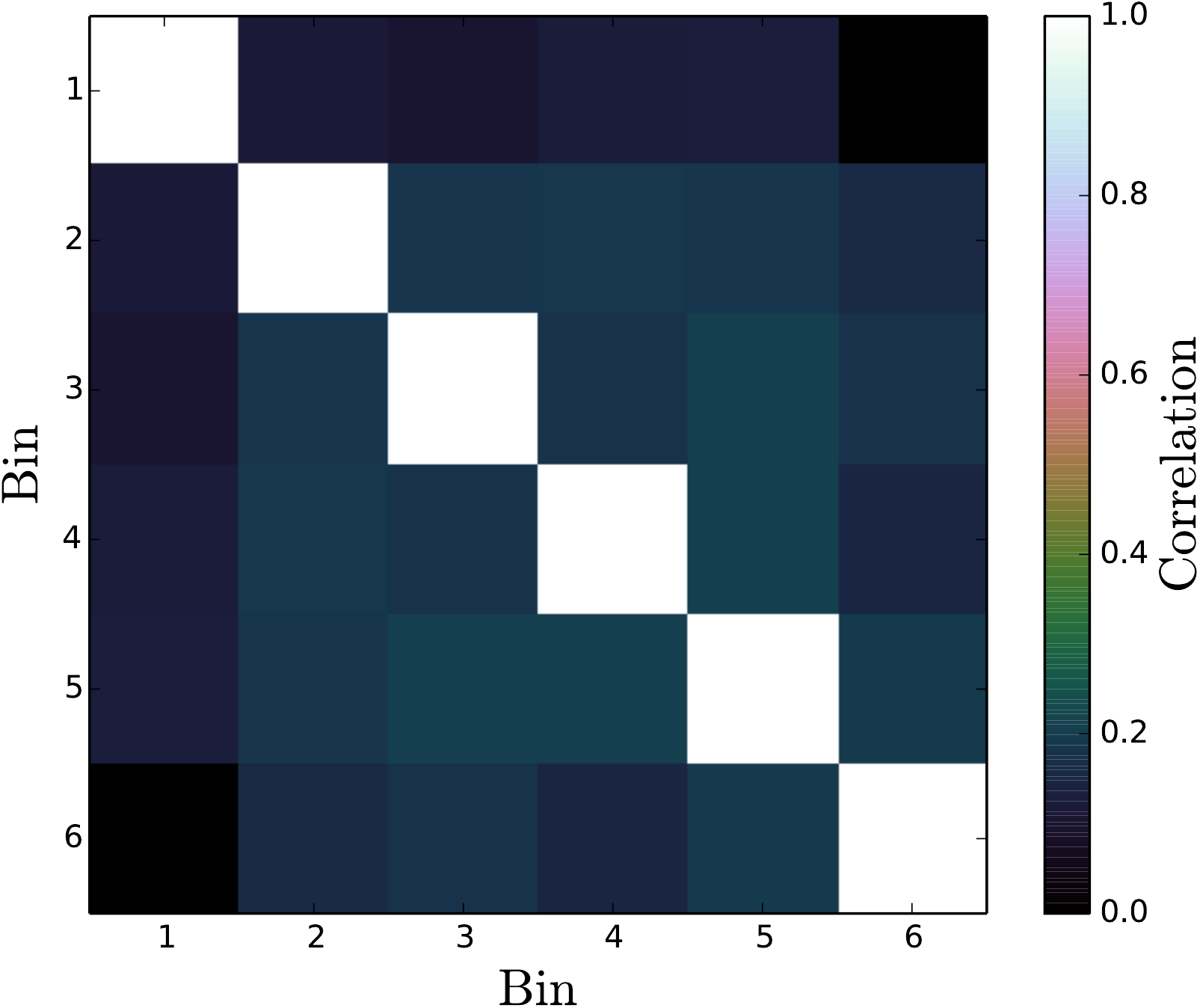}
\caption{Correlation matrix between the mean halo masses derived in the six r-band luminosity bins from the halo model fit. The reason for the correlation is the assumption of a constant scatter as a function of group luminosity in the halo occupation distribution.}
\label{fig:MassCorrelation}
\end{figure}

The inferred halo masses in each luminosity bin are slightly correlated due to the assumption that the scatter in halo mass is constant in different bins of total luminosity. 
We compute the correlation between the inferred halo masses from their posterior distribution, and we show the results in 
Figure \ref{fig:MassCorrelation}. Overall, the correlation is at most
20\%, and this is accounted for when deriving scaling relations
(see Section \ref{sec:ScalingRelation}).

\subsubsection{Concentration and mis-centring}

\begin{figure}
\includegraphics[width=8.0cm, angle=0]{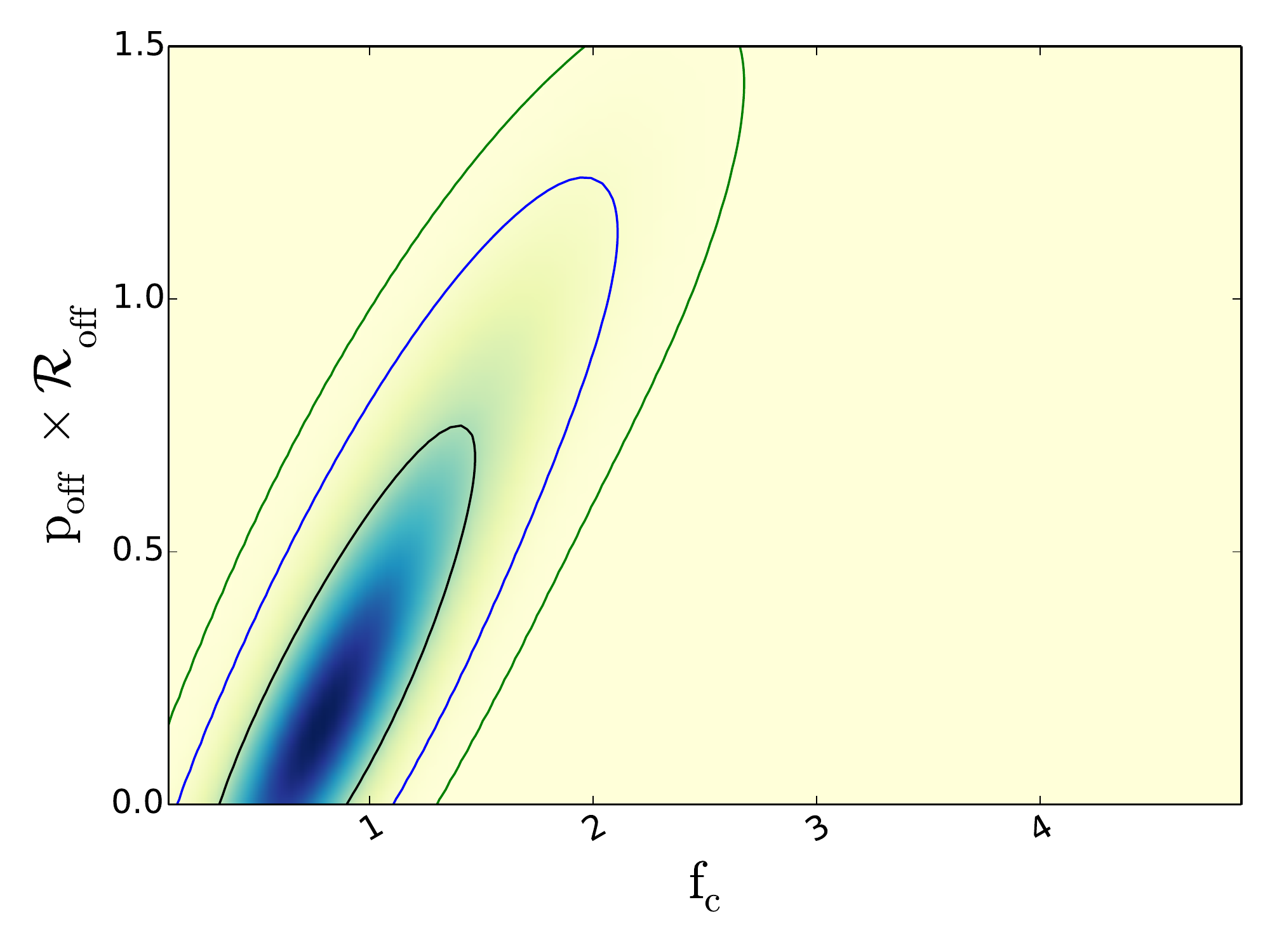}
\caption{2D posterior distribution
  of the average projected offset ($\mathrm{p_{off}} \, \times \, \cal
  R_{\rm off}$) and the normalisation of the concentration-halo mass
  relation $f_{\rm c}$. The contours indicate the 68\%, 95\% and 99\% confidence region.}
\label{fig:pr}
\end{figure}

The shape of the ESD profile at scales smaller than $\sim 200 \hk$ contains information on the concentration of the halo and on the mis-centring of the BCG with respect to the true halo centre. However, the relative normalisation of the concentration-halo mass relation, $f_{\rm c}$, and the two mis-centring parameters,
 $\mathrm{p_{off}}$ and $\cal R_{\rm off}$ are degenerate with each other. A small value of $f_{\rm c}$ has a similar effect on the stacked ESD as a large offset: both flatten the profile. To further illustrate this degeneracy, we show in Figure \ref{fig:pr} the 2D posterior distribution of the average projected offset ($\mathrm{p_{off}} \, \times \, \cal R_{\rm off}$) and the normalisation of the concentration-halo mass relation. It is clear how a vanishing offset would require a low value of the concentration.

\begin{figure}
\includegraphics[width=9cm, angle=0]{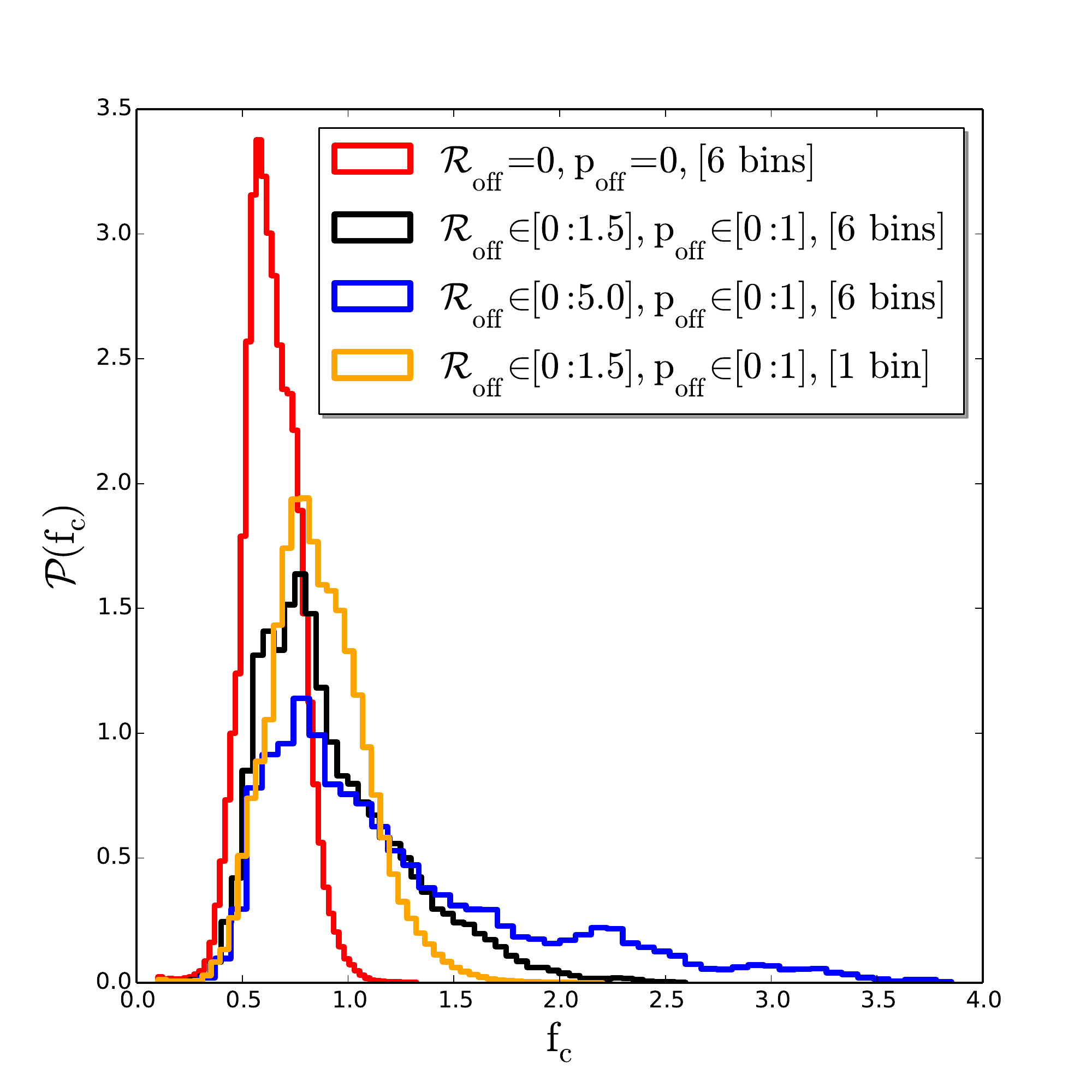}
\caption{Posterior distribution for the normalisation of the mass-concentration relation \citep{Duffy08} after marginalising over the other model parameters. We show here the effect of changing the prior in the mis-centring parameters: $\cal R_{\rm off}=\mathrm{0}$ (red line), $\cal R_{\rm off} \in \mathrm{[0..1.5]r_s}$ (black line) and $\cal R_{\rm off} \in \mathrm{[0..5]r_s}$ (blue line). As a reference, the orange line shows the posterior distribution for $f_{\rm c}$ in the case of a global stack of all groups. This has to be compared with the black line, where the constraints were derived from a joint fit of the stacked ESD in 6 luminosity bins.}
\label{fig:compFc}
\end{figure}

The derived constraints on the average projected BCG offset are quite
loose: $\mathrm{p_{off}} \, \times \, \cal R_{\rm off} < \mathrm{1.10
  r_s}$ (2-sigma). Hence one might argue in favour of a simpler model or a
model with a less informative prior on $\cal R_{\rm off}$. We address both aspects in the following ways.
First, we run a version of the halo model on the same 6  luminosity bins in which we assume no mis-centring 
(i.e. we assume that the BCG is always at the centre of the dark matter halo).  
We find a similar value of the reduced chi-squared ($\mathrm{\chi_{red}^{2}}=1.04$, 51 d.o.f.), comparable values for the 6 masses (always within one sigma) 
but tighter constraints for the relative normalisation of the concentration-halo mass relation, $f_{\rm c}=0.59^{+0.13}_{-0.11}$. 
This is perhaps not entirely surprising given that in this case  $f_{\rm c}$ is not degenerate with any other model parameter. 
Second, we relax the prior  for $\cal R_{\rm off}$ from $0\leq \cal R_{\rm off} \leq \mathrm{1.5}$ to $0\leq \cal R_{\rm off} \leq \mathrm{5}$. Also in this case, 
we find statistically equivalent halo masses and similar constraints on  $\mathrm{p_{off}}$, $\cal R_{\rm off}$, and $f_{\rm c}$ as in the fiducial case.
We summarise the results of these tests in Figure \ref{fig:compFc}.   We conclude that the fact that the reduced  $\chi^2$ values for the three model-configurations 
are very similar and always larger than unity suggests that the 11-parameter model is not too complex given the signal-to-noise of the data. 
Ignoring the mis-centring in the model lowers the relative normalisation of the concentration-halo mass relation to a 3-sigma deviation from the nominal value 
of \cite{Duffy08}. 
However, we caution the reader against over-interpreting this result
as our test shows that this is probably driven by the very
strong prior on the location of the BCGs rather than actually being a
physical property of the stacked haloes.

Lower values of the normalisation of the concentration-halo mass
relation from weak lensing analysis have been previously reported. 
For example, \cite{Mandelbaum08b} studied a sample of LRGs and MaxBCG clusters from 
SDSS and reported a 2-sigma deviation of the normalisation of the mass-concentration relation with respect to the simulation predictions. In this case, the lenses were assumed to be the true centre of the dark matter halo, and the analysis limited to scales larger than 0.5 $h^{-1}\mathrm{Mpc}$ to limit the impact of mis-centring. From a weak lensing and clustering analysis of SDSS-III CMASS galaxies, \cite{Miyatake13} also found a lower normalisation if mis-centring of the lenses is not included in the model but report agreement with the theoretical predictions once the 
mis-centring is included. A similar conclusion was derived by \cite{vUitert15} from a lensing analysis of LOWZ and CMASS LRGs from the Baryon Oscillation Spectroscopic Survey (BOSS) SDSS-DR10 using imaging data from the second Red-sequence Cluster Survey (RCS2).
In an analysis of the CFHT Stripe 82 Survey for haloes
of masses around $10^{14} h^{-1}\mathrm{M_{\sun}}$, \cite{Shan15} also reported a nominal value of the normalisation 
of the concentration-halo mass relation lower than the \cite{Duffy08} prediction, but the discrepancy between observations and predictions from
numerical simulations was not statistically significant.

Possible explanations for a lower normalisation of the concentration-halo mass relation might include halo-triaxiality, which we do not account for in our model, substructures inside the main halo \citep{Giocoli12}, galaxy formation related processes which can make halo density profiles shallower by expelling baryons into the outer region of the halo (\citealt{Sales10}, \citealt{vDaalen11}) and the assumed cosmological model. In fact, the value of the concentration at a given redshift, as a measure of the formation time of haloes, depends on the background cosmology.  To address this last point, we run the halo model assuming two alternative cosmologies: a slight deviation from the nominal Planck result $(\Omega_m, \sigma_8, h, n_s, \Omega_b h^2)=$ $(0.302, 0.818, 0.68, 0.9686, 0.02197)$ \citep{Spergel15}, and the best fit result of a clustering and lensing analysis on SDSS data $(\Omega_m, \sigma_8, h, n_s, \Omega_b h^2)=$ $(0.278, 0.763, 0.739, 0.978, 0.02279)$ \citep{Cacciato13}, which we regard as an extreme change in light of the recent Planck results. We do not find any difference in the posterior distributions of any model parameters, in particular on $f_{\rm c}$. We hence conclude that our results are not affected by the assumed cosmology. 

\subsubsection{Point mass: the innermost part of the halo}

Measurements of the ESD profile at scales smaller than $\sim 50$ $\hk$ constrain the amount of mass at the centre of the halo. We model this as a simple point mass. The measured amplitude of the point mass is not degenerate with any of the other halo model parameters, demonstrating that, given the quality of the data, the details of the distribution of the baryons at the very centre of the haloes are not relevant to infer global properties of the dark matter halo, such as their masses or concentrations.

\subsubsection{Other definitions of the group centre}
\label{sec:misCenter}

Finally, we repeat the analysis using two alternative definitions of the group centre in the GAMA catalogue: the centre of light (\texttt{Cen}) and the brighter galaxy left after iteratively removing the most distant galaxy from the group centre of light (\texttt{IterCen}). We present the results in Appendix A, Table \ref{tab:summaryParam} and \ref{tab:summaryParamExtra}.
We do not find any significant difference in the ESD profile when using \texttt{IterCen} instead of the \texttt{BCG}. 
However, the profile is very different when we use  \texttt{Cen}. In this case, we find tight constraints on the probability of the centre of light of not being the 
centre of the dark matter halo  with $\mathrm{p_{off}} \ge 0.67$ at 2-sigma and we find that on average the amount of mis-centring of the centre of light with respect to the 
minimum of the halo potential well is $\cal R_{\rm off} = \mathrm{1.00^{+0.37}_{-0.51}}$. 
The constraints on the halo masses in the 6 luminosity bins, as well as the 
constraints on $\mathrm{\sigma_{\rm log{\tilde M}}}$, 
$f_{\rm c}$, and $\mathrm{A_{P}}$, are however  consistent within 1-sigma with those calculated using the BCG position.

In summary, our results highlight the importance of a proper model for
the mis-centring in the analysis of the ESD signal from groups or
clusters of galaxies. Neglecting it could lead to biases in the derived parameters, particularly the normalisation of the concentration-mass relation.

\begin{table*}
\caption{Constraints on the average halo mass in each r-band luminosity bin using the three definitions of halo centre. We quote here the median of the mass posterior distribution, marginalised over the other halo model parameters, and the errors are the 16th and 84th percentile of the distribution. All of the constraints derived using the three different proxies for the halo centre agree within 1-sigma.}
\label{tab:summaryParam}
\resizebox{\linewidth}{!}{%
\begin{tabular}{lllllll}
\hline
$\texttt{Centre}$ & $\mathrm{\log[M_{200}^{(1)}/(h^{-1}M_{\odot})]}$&$\mathrm{\log[M_{200}^{(2)}/(h^{-1}M_{\odot})]}$ & $\mathrm{\log[M_{200}^{(3)}/(h^{-1}M_{\odot})]}$&$\mathrm{\log[M_{200}^{(4)}/(h^{-1}M_{\odot})]}$&$\mathrm{\log[M_{200}^{(5)}/(h^{-1}M_{\odot})]}$&$\mathrm{\log[M_{200}^{(6)}/(h^{-1}M_{\odot})]}$\\
\hline
\texttt{BCG} & $13.15^{+0.13}_{-0.15}$&$13.52^{+0.13}_{-0.15}$&$13.83^{+0.11}_{-0.12}$&$13.76^{+0.10}_{-0.12}$&$14.13^{+0.09}_{-0.10}$&$14.55^{+0.10}_{-0.10}$\\
\texttt{IterCen} &$13.21^{+0.12}_{-0.13}$&$13.45^{+0.13}_{-0.16}$&$13.76^{+0.11}_{-0.13}$&$13.77^{+0.10}_{-0.11}$&$14.16^{+0.08}_{-0.09}$&$14.53^{+0.09}_{-0.09}$\\
\texttt{Cen} & $13.00^{+0.17}_{-0.23}$&$13.64^{+0.12}_{-0.16}$&$13.92^{+0.10}_{-0.12}$&$13.85^{+0.10}_{-0.12}$&$14.18^{+0.09}_{-0.10}$&$14.64^{+0.10}_{-0.10}$\\
\hline
\end{tabular}}
\end{table*}

\begin{table*}
\caption{Constraints on the halo model parameters using the three definitions of halo centre. For each of the parameters, we quote the median of the posterior distribution, marginalised over the other parameters, while the errors are the 16th and 84th percentile of the distribution. All the constraints derived using the three different proxies for the halo centre agree within 1-sigma.}
\label{tab:summaryParamExtra}
\begin{tabular}{llllll}
\hline
$\texttt{Centre}$ &$\mathrm{\sigma_{\rm log[\tilde M]}}$ & $f_{\rm c}$ & $\mathrm{p_{off}}$ & $\mathrm{{\cal R}_{\rm off}}$ & $\mathrm{A_{P}}$\\
\hline
\texttt{BCG} &$0.74^{+0.09}_{-0.16}$  & $0.84^{+0.42}_{-0.23}$ & $0.38^{+0.30}_{-0.27}$ & $0.79^{+0.52}_{-0.62}$ & $2.06^{+1.19}_{-0.99}$\\
\texttt{IterCen} & $0.74^{+0.10}_{-0.16}$ & $0.94^{+0.43}_{-0.23}$ & $0.37^{+0.27}_{-0.26}$ & $0.87^{+0.46}_{-0.65}$ &$1.76^{+1.12}_{-0.87}$\\
\texttt{Cen} & $0.67^{+0.10}_{-0.17}$  & $1.10^{+0.32}_{-0.46}$ & $0.98^{+0.02}_{-0.09}$ & $1.00^{+0.37}_{-0.51}$ & $0.91^{+0.63}_{-0.33}$\\
\hline
\end{tabular}
\end{table*}

\section{Scaling relations}\label{sec:ScalingRelation}

In the last Section of this paper, we investigate the correlations
between the halo masses derived using weak gravitational lensing and
optical properties of galaxy groups measured from SDSS images and the GAMA catalogue \citepalias{Robotham11}. There are two main reasons to study these scaling relations: i) to
understand which physical processes take place inside galaxy groups
and their impact on galaxy formation; ii) to constrain a mean
relation, as well as the scatter, between some observable property of
the groups and their halo mass for use in cosmological analyses that rely on the halo mass function.

\subsection{The relation between halo mass and group r-band luminosity}

\begin{figure*}
\includegraphics[width=18cm, angle=0]{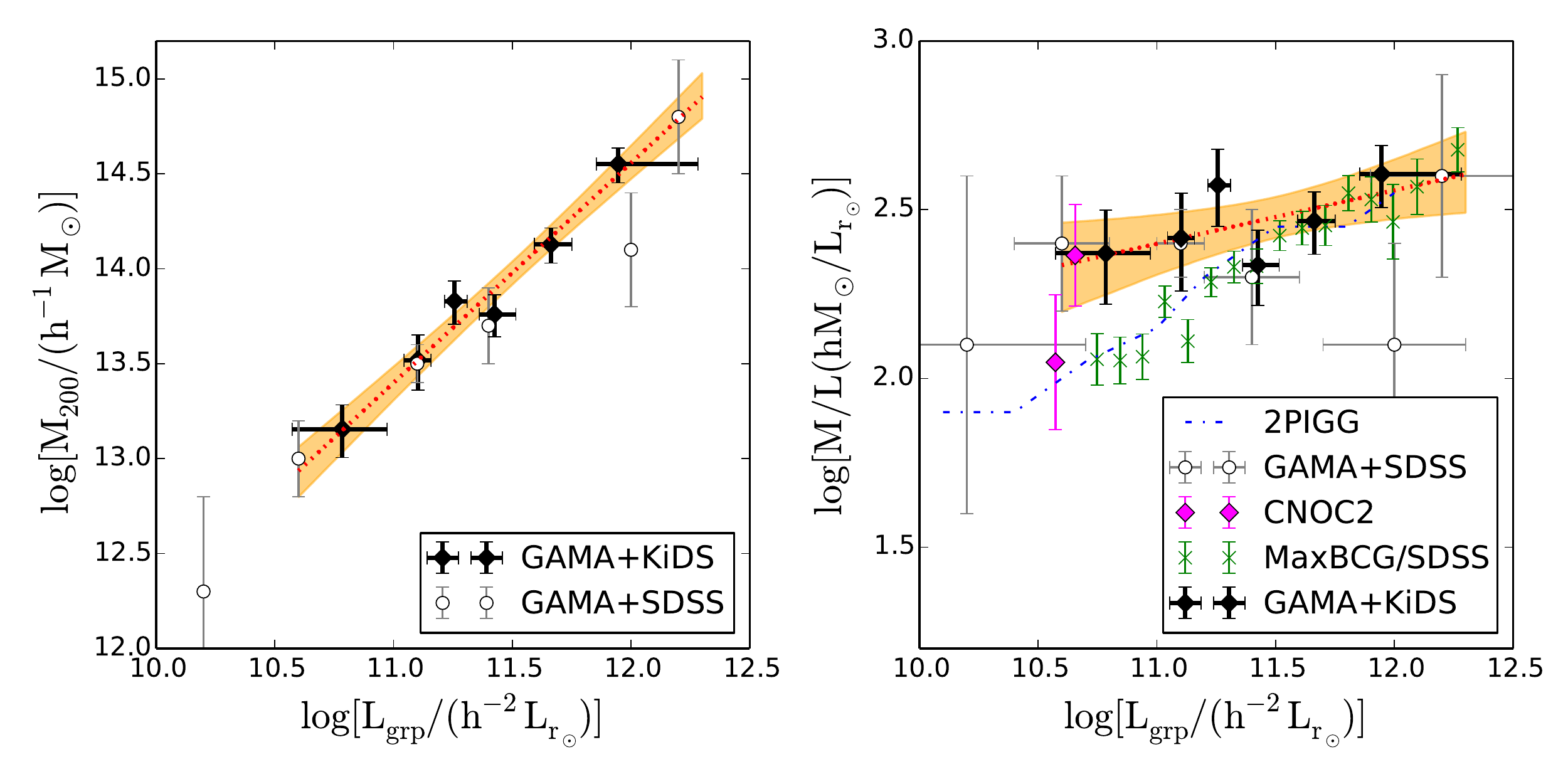}
\caption{\textit{Left panel:} Halo mass as a function of the total group r-band luminosity. The solid black points show
  the halo masses derived in this work from a halo model fit to the
  stacked ESD profile of groups with at least 5 members brighter than
  the GAMA magnitude limit. The vertical error bars indicate the 1-sigma uncertainty on the average halo mass after marginalising over the other halo model parameters, while the horizontal error bars indicate the 16th and 84th percentile of the luminosity distribution in each bin. The red line shows the best fit power-law to the data points, while our estimate of the 1-sigma {\rm dispersion} around this relation is shown as the orange area (see text). The open black circles show the halo masses derived from
  a lensing analysis of GAMA groups using SDSS galaxies as background sources \citep{Han14}. \textit{Right panel:} Derived mass-to-light ratio as a function of the group total luminosity from this work (black points), from the GAMA+SDSS analysis (open black circles), from the analysis of the CNOC2 group sample \citep{Parker05} (magenta diamonds) and from a lensing analysis of 130000 groups from the MaxBCG catalogue using SDSS imaging \citep[][(green crosses)]{Sheldon09}. In blue we show the median relation derived using the 2PIGG catalogue \citep{Eke04}. The red lines and the orange area correspond to those of the left panel.}
\label{fig:DerivedScaling}
\end{figure*}

We first investigate the scaling relation between the total halo mass and the total r-band luminosity of the groups. As described in the previous section, we bin the groups according to their total r-band luminosity (see Table \ref{tab:binSummary}), fit a halo model to the stacked ESDs, and record the halo mass posteriors for each bin. We show the results, halo mass a function of group luminosity, in the left panel of Figure \ref{fig:DerivedScaling}. 

We fit a power-law relation between the halo mass and the total r-band luminosity of the group:
\begin{equation}
\frac{\mathrm{M_{200}}}{10^{14}h^{-1}\mathrm{M_{\odot}}}=(0.95 \pm 0.14)\Bigg(\frac{\mathrm{L}_{\mathrm{grp}}}{10^{11.5} h^{-2}\mathrm{L_{\odot}}}\Bigg)^{(1.16\pm 0.13)}
\end{equation}
The linear regression is performed in the log-basis,  since the errors on the masses are log-normal distributed, by minimizing the offset of the mass measurements from the power-law relation.
We explicitly account for the correlation between halo masses (see Section \ref{sec:HaloProperies}). 
The red line in Figure \ref{fig:DerivedScaling} shows the best-fit relation. Our estimate of the 1-sigma {\rm dispersion} around this relation is shown as the orange band and is derived from the joint posterior distributions for the halo masses from 5 independent MCMCs. We jointly extract $10^{5}$ random values of the masses in each of the 6 r-band luminosity bins (in order to preserve the correlation between the masses), and we fit a linear relation to each log-mass vector as a function of the logarithm of the r-band luminosity. Finally, we compute the 16th and 84th percentiles of the best fit models in the different r-band luminosity bins. The average logarithmic scatter in halo mass at fixed r-band luminosity is $\sigma_{\log{\langle\mathrm{M_{200}}\rangle}}=0.17$

In the left panel of Figure \ref{fig:DerivedScaling}, we also compare our results to a previous weak lensing analysis of the same group catalogue (open black points)
that used SDSS galaxies as background sources \citep{Han14}. That analysis included all groups with $\mathrm{N_{fof} \ge 3}$ and fitted a single maximum 
likelihood mass to all the galaxies within a number of r-band luminosity bins.
The agreement between the two analyses is remarkable given the different quality of data and the different techniques used to infer the halo masses.  
Nevertheless, we stress that the current analysis based on the first KiDS data not only yields some of the tightest lensing constraints on group masses to date but also does this whilst marginalising over halo model parameters not considered in the previous work.  

Mock simulations suggest that the GAMA group catalogue is significantly
contaminated by chance projections for groups with 2 and 3 members and marginally contaminated for groups with 4 members. Thus, while the only way to obtain constraints on
low-luminosity systems ($\mathrm{L_{\rm grp}}\la10^{10.5}L_\odot
h^{-2}$) is to include such sparse groups in the analysis, the
impurity of the selection makes any results on the average mass of
such groups unreliable and difficult to quantify (most likely
underestimated). Our lowest-luminosity bin may suffer from
a bias due to this same richness criterion if, as seems plausible, the poorer groups that are not included at a given luminosity have systematically lower masses.

According to our current understanding of galaxy formation, one would expect the slope of the mass-luminosity relation to change towards the low-mass end, for haloes of about $10^{12}-10^{13} M_{\sun}h^{-1}$. This is mostly due to star formation being most efficient in haloes of  $\sim 10^{12}h^{-1} M_{\odot}$ (see for example \citealt{Behroozi13} and references therein), implying the dominant feedback process is mass ejection from supernovae (see e.g. \citealt{Dekel86}). However, we are only able to probe the mass-luminosity relation for haloes more massive than about $10^{13} M_{\odot}h^{-1}$. In the regime modeled here, the relation is well fitted by a single power-law.

The right panel of Figure \ref{fig:DerivedScaling} shows the relation
between halo mass and total r-band luminosity in terms of the mass-to-light ratio. The mass-to-light ratio is relatively constant with total group r-band luminosity, with a slight increase of less than 0.1 dex from the lowest to the highest luminosity bin. The scatter around this ratio is as large as 0.2 dex. Ideally, one would like to compare this result with previous results from the literature. Unfortunately, different authors use different definitions of halo masses, group luminosities are often measured in different bands, and  group selection functions might differ due to different survey depths or different algorithms used to identify groups. This might easily lead to different scaling relations, and we would like to highlight to the reader that a face-value comparison might be misleading. Despite these uncertainties, we qualitatively compare our results with previous measurements in what follows.

One of the first analyses of a large sample of groups was based on the 2dFGRS, 
using a percolation technique to identify groups while also allowing dynamical mass measurements \citep{Eke04}. 
The group luminosity was measured both in $b_{J}$ and  $r_{F}$-band. We show this result as the blue line in the right panel of Figure \ref{fig:DerivedScaling}.  We find a qualitatively similar trend of the mass-to-light ratio as a function of the total group r-band luminosity for 
$L_{\mathrm{grp}} > 10^{11}L_{r_{\odot}}h^{-2}$. However, our data do not support the steep increase of the mass-to-light ratio in the range 
$10^{10}L_{r_{\odot}}h^{-2} <L_{\mathrm{grp}}< 10^{11}L_{r_{\odot}}h^{-2}$ reported by \cite{Eke04}. 

\cite{Han14} carried out a detailed comparison between their results (which are in agreement with the one presented in this work) and the results from \cite{Eke04}, concluding that the steep increase in the mass-to-light ratio 
observed in the 2dFGRS sample could be mostly explained by the different depth between 2dFGRS and GAMA (2 magnitudes deeper).  We stress again here 
that our first data point might be affected by the apparent richness selection
we applied on the group catalogue. If we exclude this data point, the agreement with \cite{Eke04} is fairly reasonable.  

We also compare our results with a lensing analysis of MaxBCG clusters \citep{Koester07} using SDSS imaging \citep{Sheldon09}. We show their result as the green points in Figure \ref{fig:DerivedScaling}. In this case the groups/clusters were binned according to their total luminosity and the masses were measured by first inverting the ESD signal to 3D density and mass profiles and then by inferring the mass inside $R_{200}$. Also in this case we find a reasonable agreement once we exclude our first data point, which, as discussed, might be affected by the apparent richness selection we applied to the group catalogue.

Finally \cite{Parker05} considered a sample of 116 groups from the CNOC2 survey \citep{Yee98}. The halo masses 
were measured by fitting a SIS profile to the stacked ESD signal measured using weak gravitational lensing. 
In this case, the luminosity was measured in $B$-band. Given the small sample of groups, only two measurements were possible at quite low 
group luminosity. We show their results as the magenta points in Figure \ref{fig:DerivedScaling}.  Following \citet{Jee14}, we applied a 0.8 multiplicative correction to  the $B$-band mass-to-light ratio in order to have an estimate for the mass-to-light ratio in $r$-band. 
 
Only the mass-to-light ratio measurement in the high luminosity bin of the CNOC2 analysis, which corresponds to our low luminosity bin, can be directly compared to our analysis, given the luminosity range we probe. We find a good agreement. 

\subsection{The relation between halo mass and velocity dispersion}

\begin{figure}
\includegraphics[width=8.5cm, angle=0]{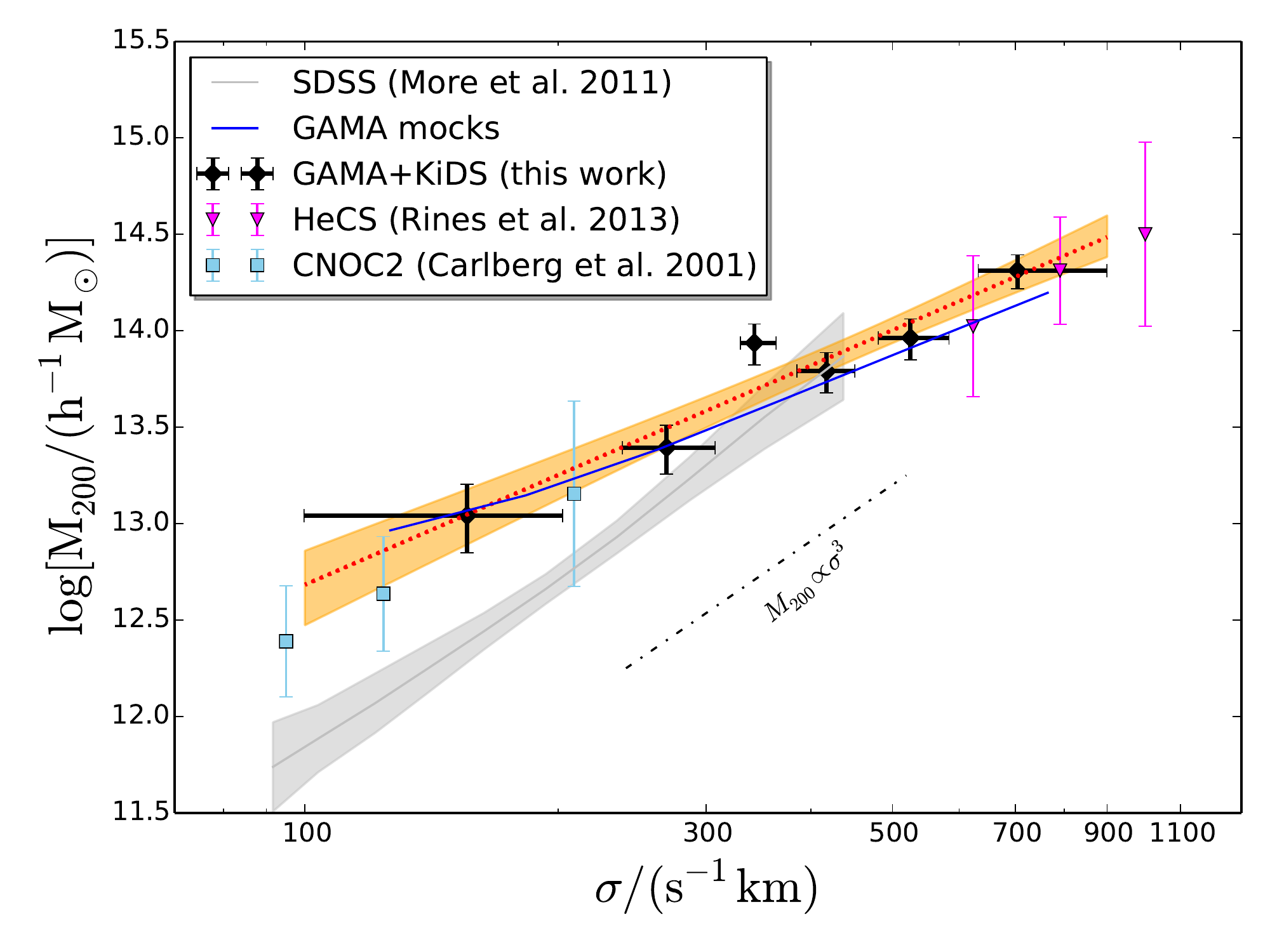}
\caption{Halo mass as a function of the
  group velocity dispersion. The black points show the halo masses
  derived in this work from a halo model fit to the stacked ESD
  profile of groups with at least 5 members brighter than the GAMA
  magnitude limits. The red line shows the best fit power-law to the data
  points and the orange area
  indicates our estimate of the 1-sigma dispersion around this
  relation. The cyan points show the results from the CNOC2 survey
  \citep{Carlberg01}, while the magenta points show the results from
  the HeCS sample of clusters \citep{Rines13}. The grey band shows the
  mass-velocity dispersion relation obtained from measurements of
  satellite kinematics in SDSS \citep{More11}. Finally, the blue line shows the relation calculated from the GAMA mocks using the same selection function applied to the data.}
\label{fig:DerivedScalingVel}
\end{figure}

Next, we focus on the scaling relation between the total halo mass and the group velocity dispersion. 
Again, we bin the groups in 6 bins  according to their velocity dispersion, with the boundaries chosen so that the signal-to-noise ratios of the stacked ESD profiles are equal (see Table  ~\ref{tab:binSummary}).
The halo masses in each bin are then found by a joint halo model fit to the ESD profile in each velocity dispersion bin. Figure \ref{fig:DerivedScalingVel} shows the corresponding results. The GAMA groups span an order of magnitude in velocity dispersion, but most of the constraining power for the scaling relation comes from groups with $\mathrm{\sigma} \sim 500 \mathrm{km\,s}^{-1}$. This is expected given that the cut imposed on group apparent richness excludes the low mass systems from this analysis, and that the survey volume is relatively small, and hence our sample does not contain many very massive galaxy clusters.
As in the case of binning by luminosity, we believe that the apparent richness cut imposed on the GAMA group catalogue will have a non-neglible effect on the measurement of the average halo mass in the first velocity dispersion bin $\sigma < 200 \mathrm{km\,s}^{-1}$.

At low velocity dispersion, we compare our results with those from the
CNOC2 survey \citep{Carlberg01}, for which the mass measurements are derived from the dynamical properties of the groups. In Figure \ref{fig:DerivedScalingVel} we show the average CNOC2 mass measurements in 3 velocity dispersion bins; the error bars are the 1-sigma scatter between measurements in each bin. 

At high velocity dispersion, we compare our results to the analysis of the HeCS sample \citep{Rines13}, where masses are measured using a redshift-space caustic technique. The mean redshift of the HeCS clusters is similar to that of the GAMA groups. As for the CNOC2 sample, 
we binned the HeCS clusters according to their velocity dispersion, and we calculated the median mass and the 1-sigma dispersion in each bin. 
Both the CNOC2 and the HeCS sample agree well with the mass-velocity dispersion relation we derived using galaxy groups from GAMA. 

We fit a power-law between the halo mass and the group velocity dispersion (using the same procedure outlined in the previous section) and we constrain this relation to be:

\begin{equation}
\Bigg(\frac{M_{200}}{10^{14}h^{-1}\mathrm{M_{\odot}}}\Bigg)=(1.00 \pm 0.15)\Bigg(\frac{\sigma}{500 \mathrm{s^{-1}km}}\Bigg)^{(1.89\pm 0.27)} \, ,
\end{equation}
We find that the average scatter in the halo mass-velocity dispersion relation is $\sigma_{\log \langle \mathrm{M_{200}} \rangle}=0.20$.

We do not see any indication of a change in the slope over almost two order of magnitude in mass, from massive clusters to small groups. However, the slope we find is significantly shallower than what would be expected from a virial scaling relation ($M \propto \sigma^3$) as is seen in dissipationless numerical simulation \citep{Evrard08}. A very similar result ($M \propto \sigma^{2.09 \pm 0.34}$) was found by a previous weak lensing analysis of the same group catalogue using SDSS galaxies as background sources \citep{Han14}.

There are at least two possible explanations for this effect: 
\begin{itemize}
\item Hydrodynamical simulations have shown that galaxies trace shallower mass-velocity dispersion relations (slope lower than 3) than dark matter particles \citep{Munari13}. This is due to dynamical friction and tidal disruption, acting on substructures and galaxies, but not on dark matter particles. The typical effect measured in simulations is of order 10\%, which is too small to explain the value of the power-law slope we measure when comparing with the virial expectation. 
\item The apparent richness cut we imposed to the group catalogue,  the GAMA selection function and the limited cosmological volume we probe might introduce selection biases on our mass measurements. In particular the apparent richness cut might introduce a positive bias for mass-measurements in the lowest velocity dispersion bin, and the small volume used in this work might introduce negative biases in the highest velocity dispersion bins. The combination of these two effects would result in a shallower mass-velocity dispersion relation.  
\end{itemize} 
To investigate the second hypothesis further we compare our inferred scaling relation with one measured from the dark matter only mock GAMA catalogue \citep{Robotham11,Merson13} applying the same apparent richness cut. In the GAMA mocks the velocity dispersion is measured using the underlying/true dark matter haloes while the stored mass of the haloes (DHalo mass) are computed as the sum of the masses of their component subhaloes \citep{Jiang14}. For the purpose of the comparison we convert them into $M_{200}$ (McNaught-Roberts in prep.). We show the results as the blue line in Figure \ref{fig:DerivedScalingVel}. We find a good agreement with the scaling relation measured from the data, supporting the hypothesis that the shallower scaling relation we measure is mostly caused by selection effects. However we cannot exclude at this stage that part of the reason for the shallower mass-velocity dispersion relation might be dynamical processes acting on the galaxies in the groups.

A detailed investigation will be presented in a forthcoming paper (Robotham et al. in prep.) in the context of finding optimal dynamical mass estimates using weak lensing measurements of the group masses.

Finally, we compare our results with measurements of the mass-velocity
dispersion relation obtained from measurements of satellite kinematics
in SDSS \citep{More11}. In this case, we extrapolate the mass-velocity
dispersion relation from measurements of the stellar mass - halo mass
and stellar mass - velocity dispersion relations which are provided in
that paper. Note that these two relations have not been derived
independently from each other. We find a good agreement with our
results for $\sigma > 300 {\mathrm{km}\,\mathrm{s}^{-1}}$. For lower mass haloes, we have already discussed
the potential selection effect due to the apparent richness cut that affects our
first data point. However, we also note that there is some tension
between the CNOC2 results \citep{Carlberg01} and the SDSS satellite
kinematics results. In general, velocity dispersion and
mass measurements are more difficult for low mass groups than for
massive systems because of the smaller number of members and more
severe selection effects.

\subsection{The relation between halo mass and r-band luminosity fraction of the BCG}\label{sec:ML}

\begin{figure}
\includegraphics[width=8.5cm, angle=0]{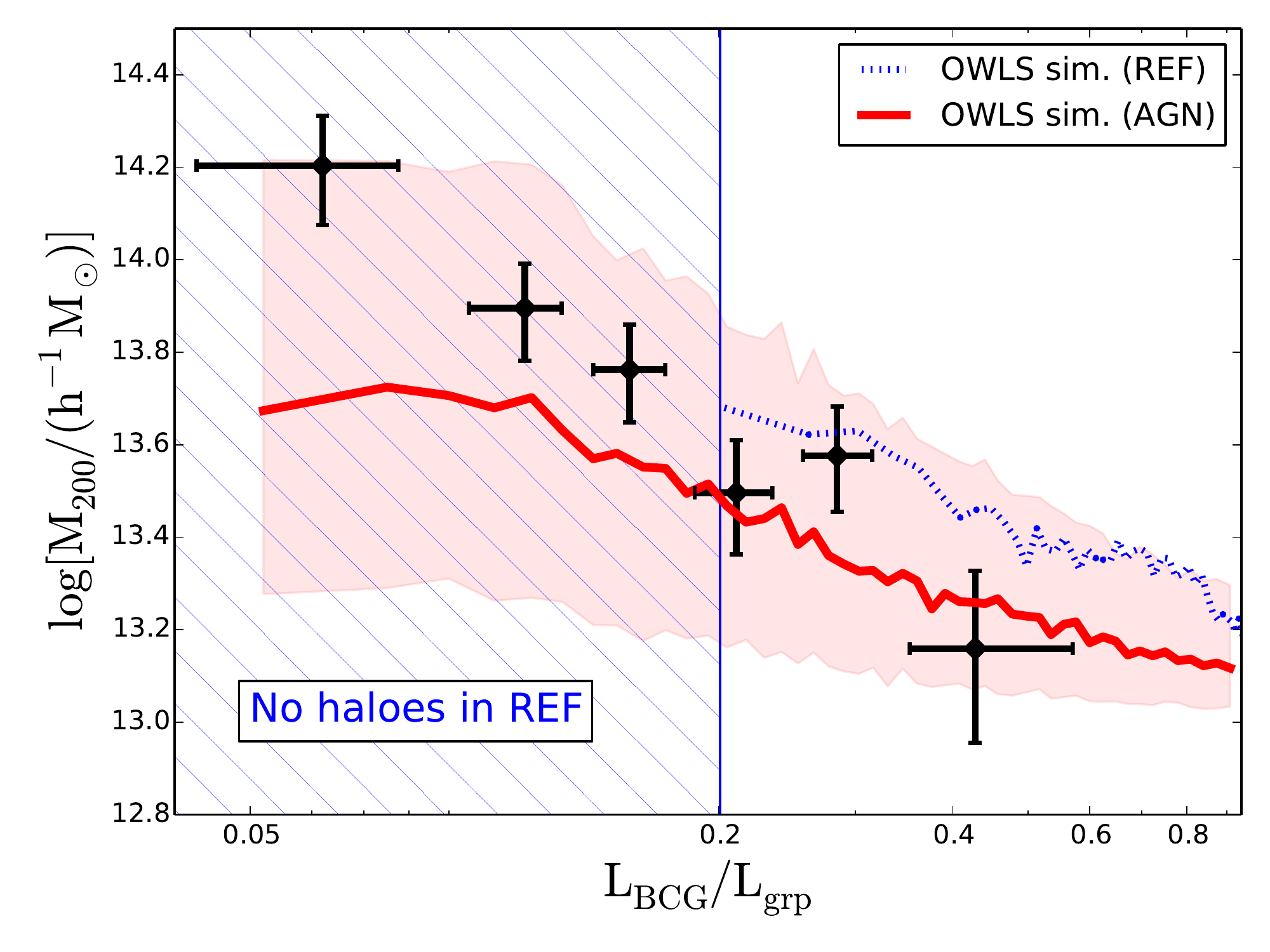}
\caption{Group masses as a function of the $r$-band luminosity fraction of the BCG. The solid black points show the halo masses derived in this work from a halo model fit to the stacked ESD profile of groups with at least 5 members brighter than the GAMA magnitude limit. The solid red and the dashed blue lines are predictions from the Cosmo-OWLS simulation at the median redshift of the GAMA groups  for a run including AGN feedback and a reference run without AGN feedback \citep{LeBrun14}. The luminosities measured in the simulation are $(k+e)$ corrected to redshift 0 using the same functional form (Equation \ref{eq:kpe}) applied to the data. The red area encompasses the 16th and 84th percentile of the mass distribution in each luminosity fraction bin for the AGN simulation. The shaded blue area indicates the range in $L_{\mathrm{BCG}}/L_{\mathrm{grp}}$ in which there are no haloes in the REF simulation.}
\label{fig:FracLightBCG}
\end{figure}

Feedback from supernovae \citep{Dekel86} and AGNs \citep{Springel05b}
have been proposed in the past decade as a possible solution for
reducing the star formation efficiency in hydrodynamical simulations
(e.g. \citealt{Sijacki07}, \citealt{Fabjan10}, \citealt{McCarthy10},
\citealt{Booth13}, \citealt{Vogelsberger14}, \citealt{Schaye15} and references therein). It is important to test the hypothesis of feedback and to constrain its
efficiency by comparing complementary predictions of hydrodynamical
simulations with observations. Motivated by the work of
\cite{LeBrun14}, we focus here on the relation between the $r$-band
luminosity fraction of the BCG, defined as $\mathrm{L_{BCG}/L_{grp}}$,  and the
group halo masses calculated in this work by binning the groups according to $\mathrm{L_{BCG}/L_{grp}}$ (see Table  ~\ref{tab:binSummary}). The $r$-band luminosity of
the BCG is calculated from the $\mathrm{r_{AB}}$ petrosian magnitude
from the GAMA catalogue. We apply a k-correction and evolution correction to the magnitude following \citetalias{Robotham11}:
\begin{equation}
\label{eq:kpe}
{\mathrm(k+e)}(z)=\sum_{i=0}^{4}a_{i}(z-0.2)^{i}-1.75z \, ,
\end{equation}
with $a_{i}=[0.2085,1.0226,0.5237,3.5902,2.3843]$. We note that the original correction presented in Equation 8 in \citetalias{Robotham11} presents an error in the sign of the last term in the above equation. Figure \ref{fig:FracLightBCG} shows the halo masses obtained for groups stacked according to $\mathrm{L_{BCG}/L_{grp}}$ as a function of $\mathrm{L_{BCG}/L_{grp}}$.

There is a clear trend of group masses with the r-band luminosity fraction of the BCG. This trend has been previously observed at group scales by \cite{Rasmussen09} and at cluster scales by \cite{Lin04}. The explanation is that the growth of the BCG is modest compared with the growth of the entire group. 

Since the luminosity of the BCG is proportional to its stellar mass content and the group luminosity is an increasing function of the total halo mass, one can compare the results reported in this paper with studies of the stellar-to-halo mass ratio (SHMR) as a function of halo mass  \citep[e.g.][]{George11, Leauthaud12b, vdBurg14, Coupon15}. There is a clear consensus on the decline of the SHMR with halo mass, which is a different manifestation of the trend dispayed in Figure \ref{fig:FracLightBCG} where we report the halo mass as a function of the r-band luminosity fraction of the BCG. In particular, for central galaxies, it has been shown \citep{Behroozi13, Coupon15} that halos of $\sim 2\times 10^{14} h^{-1}\mathrm{M_{\odot}}$ have a SHMR about an order of magnitude lower than that of halos of  $\sim 10^{13} h^{-1}\mathrm{M_{\odot}}$, again in qualitative agreement with the result shown in Figure \ref{fig:FracLightBCG}. 
The steep decline of the relation between the group mass and the r-band luminosity fraction is a consequence of star formation becoming less efficient in more massive halos. 
Several mechanisms, beyond AGN feedback, have been invoked to explain this phenomenon such as halo mass quenching \citep[e.g][]{Peng10,Ilbert13} or the presence of many satellite galaxies in massive halos which cut off the gas supply to the BCG \citep{Aragon14}.

We focus here in particular on comparing our results with the (Cosmo-) OverWhelmingly Large Simulations (OWLS) \citep{Schaye10,LeBrun14}.

\cite{LeBrun14} present results from these simulations in terms of $K$-band luminosity binned by halo mass. They report a very similar trend to the
one we observe in our data. In particular, they find a large difference in the luminosity fraction of the BCG when they compare simulations with and without AGN feedback. To compare our results with the Cosmo-OWLS simulation, the r-band results were provided by the Cosmo-OWLS team using the same K-correction and evolution correction we applied to the data (Equation \ref{eq:kpe}) for three redshifts snapshots $z=[0.125,0.25,0.375]$. When comparing the simulations to the data, we use the results from the snapshots closer to the median redshift of the GAMA groups.  We discarded from the simulation all haloes with mass lower than $10^{13} h^{-1}M_{\odot}$, which roughly corresponds to the minimum mass of groups with more than 5 members in the G$^3$Cv7 catalogue (see section ~\ref{sec:ScalingRelationNfof}). In this way we try to mimic the selection we applied to the data. Finally, we bin the simulation in the same way we bin the data, using the BCG luminosity fraction as a proxy for the group mass. 

Figure \ref{fig:FracLightBCG} shows the Cosmo-OWLS results for the run including AGN feedback (solid red line) and for a reference run (REF) without AGN (dashed blue lines). The red area encompasses the 16th and 84th percentile of the mass distribution in each luminosity fraction bin. 

For $\mathrm{L_{BCG}/L_{grp}} < 0.2$ the reference run does
  not contain any groups which on the contrary are clearly present in
  our group sample. The reason for this is that the gas cooling in the REF simulation is too efficient, leading to BCGs which are always very luminous in comparison to the total luminosity of the group. This evidence alone is sufficient to conclude
that the data disfavour a model without AGN feedback. Note that this
conclusion is independent of the group mass measurements. Our derived scaling relation between the halo mass and the luminosity fraction of the BCG for  $\mathrm{L_{BCG}/L_{grp}} > 0.2$ further supports the above conclusion, being in reasonable agreement with the prediction from the simulation including AGN feedback.
A detailed comparison of the trend in Figure~\ref{fig:FracLightBCG} with
simulations would require replicating the GAMA group finder and
selection function on the Cosmo-OWLS simulations and is beyond the scope of this paper. 

\subsection{The relation between halo mass and group apparent richness}
\label{sec:ScalingRelationNfof}

\begin{figure}
\includegraphics[width=8.5cm, angle=0]{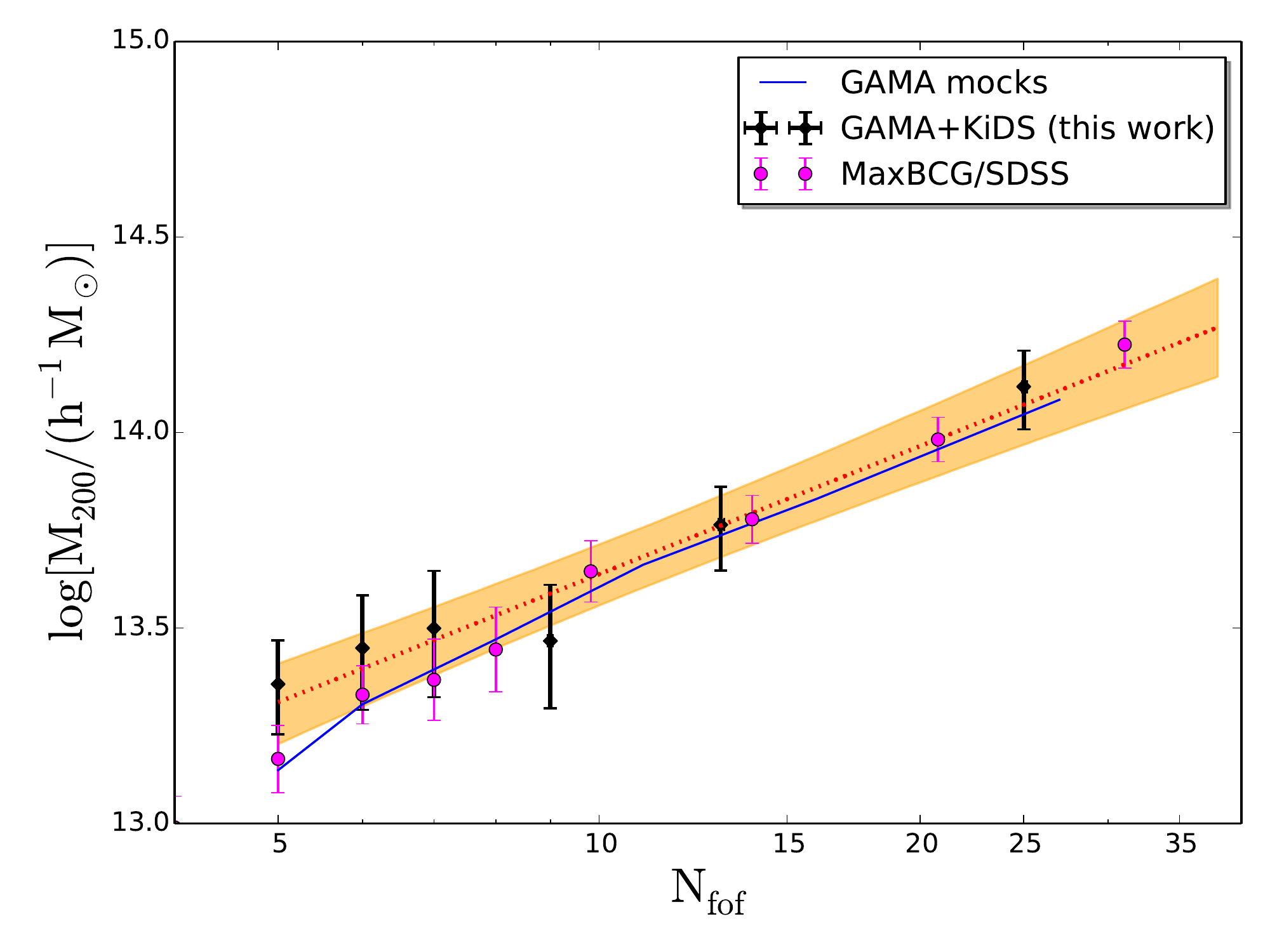}
\caption{Group halo mass as a function of the richness ($\mathrm{N_{fof}}$) . We use here only groups with at least 5 members brighter than $r_{AB}=19.8$. The richness of the groups is not corrected to account for the fainter galaxies not targeted by GAMA. The red line shows the best fit power-law relation to the data points. Our estimate of the 1-sigma dispersion around this relation is shown as the orange area. The blue line shows the mass-richness relation derived from the GAMA mocks using the same selection function applied to the data. The magenta points show the result of a weak lensing analysis of 130,000 groups and clusters of galaxies in the Sloan Digital Sky Survey \citep{Johnston07}.}
\label{fig:Nfof_scaling}
\end{figure}

Finally, we investigate the relation between the total halo mass and the apparent richness of the groups. The groups are binned according to their apparent richness (see Table  \ref{tab:binSummary}), and the average halo mass for each bin is estimated by fitting a halo model to the stacked ESD profile. We show the result in Figure \ref{fig:Nfof_scaling}. 

We parametrise the halo mass-richness relation with a power-law, which is fit to the data with the same procedure outlined in the previous sections:
\begin{equation}
\Bigg(\frac{M_{200}}{10^{14}h^{-1}\mathrm{M_{\odot}}}\Bigg)=(0.43 \pm 0.08)\Bigg(\frac{N_\mathrm{fof}}{10}\Bigg)^{(1.09 \pm 0.18)}\, ,
\end{equation}
and we constrain the average scatter in the halo mass-richness relation to be $\sigma_{\log \langle \mathrm{M_{200}} \rangle}=0.20$.

As expected, richer groups are also more massive. 
We caution the reader that this scaling relation is the one most affected by the GAMA selection function. In fact, unlike our treatment of the total group luminosity, we do not correct the apparent richness measurements to account for the faint galaxy members not targeted by GAMA.
We compare our results with the GAMA mocks, which have the same selection function as the data, and we generally find good agreement. 

We also compare our results with a weak lensing analysis of 130,000 groups and clusters of galaxies in the Sloan Digital Sky Survey \citep{Johnston07}. The masses were derived from fitting an halo model to the stacked ESD profile in 12 richness bins. The richness was defined as the number of red sequence galaxies with luminosities larger than $0.4\mathrm{L_{\star}}$ within a given projected radius, which is close to $R_{200}$. In spite of the different richness defintions we find a good agreement with our measurements, both for the amplitude and the slope of the mass-richness relation.

\section{Conclusions}\label{sec:Conclusions}

In this paper, we present the first weak lensing analysis of the mass distribution in the GAMA groups using background sources from the overlapping KiDS survey. The effective overlapping area (accounting for masks) used in this work is 68.5 square degrees and corresponds to the first two data releases of $ugri$ images of the KiDS data (\citealt{dejong/etal:2015} and \citetalias{kuijken/etal:2015}).

Our main results are the following:
\begin{enumerate}
\item
We measure the stacked excess surface density profile of the galaxy groups as a function of their total r-band luminosity, velocity dispersion, fraction of group light in the BCG and apparent richness.
Splitting the data into six roughly equal signal-to-noise bins, we derive average halo masses per bin with a typical precision of 0.12 dex. 
We provide a physical interpretation of the signal using the halo model. 
\item
We show the importance of modelling the mis-centring of the BCG (used here as tracer of the group centre) with respect to the centre of the group's dark matter halo in order to derive unbiased results, in particular on the halo mass-concentration relation.
\item
Our results are consistent with the normalisation of the halo mass-concentration relation proposed by \cite{Duffy08},when mis-centring is included in the model.
\item
We find no evidence of a significant baryonic component in the centre
of the groups in excess of the stellar mass of the BCG. However, the
uncertainty on this result is quite large due to the low
signal-to-noise at small scales, which is in turn caused by the difficulties inherent in measuring reliable shapes for blended objects.
\item
We obtain clear scaling relations between the halo mass and a number
of observable properties of the groups: the group r-band luminosity, the velocity dispersion of the group, its apparent richness and the ratio between the r-band luminosity of the BCG and the total r-band luminosity of the group. The typical scatter in halo mass at fixed observable property is $\sigma_{\log \langle \mathrm{M_{\rm 200}} \rangle}=0.2$.
\item  We show that our data have the statistical power to discriminate between models with and without  AGN feedback and possibly between different AGN feedback models.
\end{enumerate}

This analysis is part of the first set of weak lensing results using the KiDS data, based on data obtained during the first two years of operation. As the survey continues to cover more sky, both the statistical power and the fidelity of the measurements will grow, further refining these results as well as enabling other analyses of the distribution of dark matter in galaxies, groups and clusters.

\section*{Acknowledgements}

We would like to thank the anonymous referee for providing useful suggestions to improve the manuscript. 
We thank Tom Kitching, Joachim Harnois-Deraps, Martin Eriksen and Mario Radovich for providing useful comments to the paper. We are grateful to Matthias Bartelmann for being our external blinder, revealing which of the four catalogues analysed was the true unblinded catalogue at the end of this study. We would like to thank Tamsyn Mcnaught-Roberts for providing the conversion between the halo mass definition used in the GAMA mocks and $M_{200}$, and Ian McCarthy for providing the data used in Figure 15. We also thank Ludo van Waerbeke for writing the code used to compute the shear additive bias correction used in this work. MV, MC, H.Ho, CS, AC and CH acknowledge support from the European Research Council under FP7 grant number 279396 (MV, MC, CS, H.Ho) and 240185 (AC and CH). BJ acknowledges support by an STFC Ernest Rutherford Fellowship, grant reference ST/J004421/1. EvU acknowledges support from an STFC Ernest Rutherford Research Grant, grant reference ST/L00285X/1. PN acknowledges the support of the Royal Society through the award of a University Research Fellowship, the European Research Council, through receipt of a Starting Grant (DEGAS-259586) and 
support of the Science and Technology Facilities Council (ST/L00075X/1). RN and EvU acknowledge support from the German Federal Ministry for Economic Affairs and Energy (BMWi) provided via DLR under project no.50QE1103. H.Hi. is supported by the DFG Emmy Noether grant Hi 1495/2-1. 
This work is supported by the Netherlands Organisation for Scientific Research (NWO) through grants 614.001.103 (MV) and 614.061.610 (JdJ) and  by the Deutsche Forschungsgemeinschaft in the framework of the TR33 'The Dark Universe'. This work is based on data products from observations made with ESO Telescopes at the La Silla Paranal Observatory under programme IDs 177.A-3016, 177.A-3017 and 177.A-3018.
GAMA is a joint European-Australasian project based around a spectroscopic campaign using the Anglo-Australian Telescope. The GAMA input catalogue is based on data taken from the Sloan Digital Sky Survey and the UKIRT Infrared Deep Sky Survey. Complementary imaging of the GAMA regions is being obtained by a number of independent survey programs including GALEX MIS, VST KiDS, VISTA VIKING, WISE, Herschel-ATLAS, GMRT and ASKAP providing UV to radio coverage. GAMA is funded by the STFC (UK), the ARC (Australia), the AAO, and the participating institutions. The GAMA website is http://www.gama-survey.org/.

{\it Author Contributions}:  All authors contributed to the development and writing of this paper.  The authorship list reflects the lead authors (MV, MC, MB, KK) followed by two alphabetical groups.  The first alphabetical group includes those who are key contributors to both the scientific analysis and the data products.  The second group covers those who have either made a significant contribution to the data products, or to the scientific analysis.
\appendix

\section{Alternative definitions of the group centre}
\label{sec:OtherCentre}

We present here the measurements of the stacked ESD profile and the halo model constraints we obtain if we use a different definition of the group centre, compared to the BCG definitions used throughout the paper. 

\begin{figure*}
\includegraphics[width=18cm, angle=0]{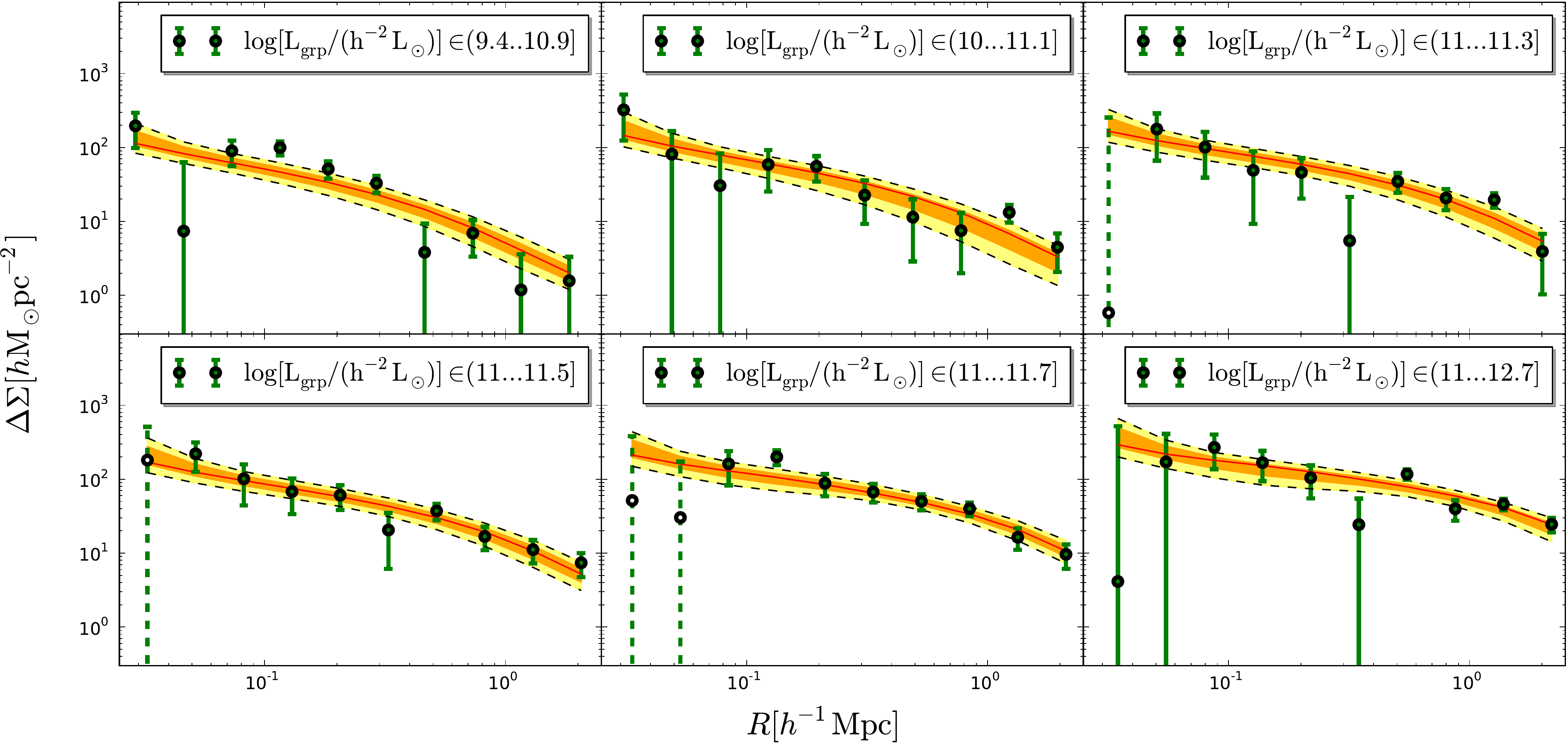}
\includegraphics[width=18cm, angle=0]{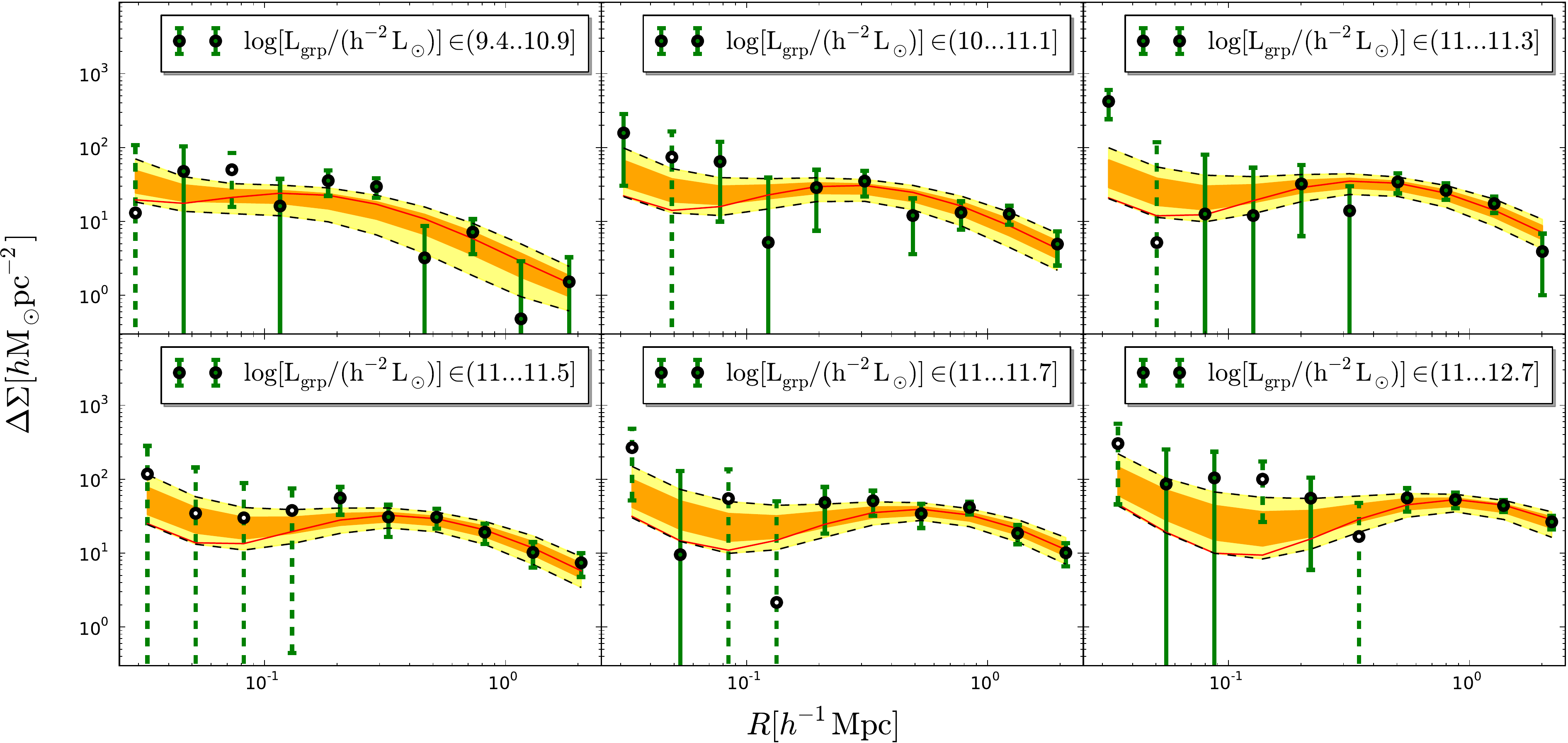}
\caption{Stacked ESD measured around the groups' \texttt{IterCen} (\textbf{upper panel}) and the groups' centre of light \texttt{Cen} (\textbf{lower panel}) for 6 group luminosity bins as a function of distance from the group centre. The group luminosity increases from left to right and from top to bottom. The stacking of the signal has been performed considering only groups with $\mathrm{N_{fof}} \ge 5$. The error bars on the stacked signal are computed as detailed in section \ref{sec:error}. The orange and yellow bands represent the 68 and 95 percentile of the model around the median and the red lines indicate the best fit model.}
\label{fig:GroupSignalOtherCentre}
\end{figure*}

In Figure \ref{fig:GroupSignalOtherCentre} we show the stacked ESD
profile for the same 6 luminosity bins used in
Section \ref{sec:HaloProperies} but now using the brightest galaxy left after iteratively removing the most distant galaxies from the group centre of light which is labelled as \texttt{IterCen} (left
panel) and the group centre of light \texttt{Cen} (right panel) as the definition for the group centre. 
When \texttt{IterCen} is used, the stacked signal is statistically
indistinguishable from the case when \texttt{BCG} is used as the group
centre. This is not surprising since the two centre definitions differ
only for a few percent of the groups.
 
When \texttt{Cen} is used as the group centre, the shape of the
stacked ESD profile is very different. The turnover of the signal at scales around 100 $h^{-1}\mathrm{kpc}$ is a clear indication of mis-centring between the chosen centre of the halo group and the true minimum of the halo potential well. \citetalias{Robotham11} report that \texttt{Cen} is not a good proxy for the halo centre, and hence, this result is not surprising. 
It is clear in this case that not including the mis-centring parameters in the model would lead to a very poor description of the data.

We do not show the posterior distributions
for the halo model parameters corresponding to the case of
\texttt{Cen}. The degeneracies between the parameters are the same as
those found when \texttt{BCG} or \texttt{IterCen} are used as proxies for the halo centre.
We can derive tight constraints on the probability of mis-centring $\mathrm{p_{off}} \ge 0.67$ 2-sigma, and we find that on average the amount of offset of the centre of light with respect to the minimum of the halo potential well is $\cal R_{\rm off} =\mathrm{1.00^{+0.37}_{-0.51}}$. 
We summarise the results in Table ~\ref{tab:summaryParam} and ~\ref{tab:summaryParamExtra}.

The constraints we derive for the halo masses in the 6 luminosity bins and the constraints on $\mathrm{\sigma_{\rm log{\tilde M}}}$, $f_{\rm c}$, $\mathrm{A_{P}}$ are consistent within 1-sigma with the constraints derived using the other two definitions of the halo centre.  

These results highlight the importance of a proper model of the
mis-centring in the analysis of the lensing signal from groups or clusters of galaxies. Neglecting mis-centring could lead to biases in the derived masses and in the other model parameters, particularly the halo concentration.
\newline
\bibliographystyle{mn2e_v1}
\bibliography{biblio}

\end{document}